\documentclass[
	acmsmall,
	screen,
	nonacm
]{acmart}
\usepackage{pifont,multirow,stfloats,makecell}
\usepackage{circledsteps}
\usepackage{amsmath}
\usepackage[inline]{enumitem}
\usepackage{graphicx}
\usepackage[singlelinecheck=false]{caption}
\usepackage{subcaption}
\usepackage{wrapfig}
\usepackage{tikz}
\usetikzlibrary{shapes, arrows, positioning}
\usepackage{pdflscape}
\usepackage{tablefootnote}
\usepackage{afterpage}
\usepackage[para]{footmisc}

\newcommand\bcircle[2][]{\ifmmode
\Circled[fill color=white,inner color=black,#1]{\mathsf{#2}}
\else
\Circled[fill color=white,inner color=black,#1]{\sffamily#2}
\fi
}

\PassOptionsToPackage{destlabel}{hyperref}

\newenvironment{wrapped}[1]
	{\def\wrappedcurrent{#1}
		\setlength{\columnwidth}{\parshapelength\numexpr\prevgraf+2\relax}
		\csname #1\endcsname}
	{\csname end\wrappedcurrent\endcsname}

\setcopyright{none}
\acmDOI{}
\acmISBN{}
\acmJournal{PACMPL}
\acmYear{-}
\acmMonth{-}
\acmVolume{-}
\acmVolume{-}
\acmArticle{-}
\copyrightyear{}

\begin{document}
\title{Solving Package Management via Hypergraph Dependency Resolution}

\author{Ryan Gibb}
\orcid{0009-0009-5702-3143}
\email{ryan.gibb@cl.cam.ac.uk}
\affiliation{
	\institution{University of Cambridge}
	\country{United Kingdom}
}

\author{Patrick Ferris}
\orcid{0000-0002-0778-8828}
\affiliation{
	\institution{University of Cambridge}
	\country{United Kingdom}
}
% \email{patrick.ferris@cl.cam.ac.uk}

\author{David Allsopp}
% \orcid{}
\affiliation{
	\institution{Tarides}
	\country{United Kingdom}
}
% \email{david@tarides.com}

\author{Michael Winston Dales}
% \orcid{}
\affiliation{
	\institution{University of Cambridge}
	\country{United Kingdom}
}
% \email{mwd24@cam.ac.uk}

\author{Mark Elvers}
% \orcid{}
\affiliation{
	\institution{Tarides}
	\country{United Kingdom}
}
\email{mte24@cam.ac.uk}

\author{Thomas Gazagnaire}
% \orcid{}
\affiliation{
	\institution{Tarides}
	\country{France}
}
\email{thomas@tarides.com}

\author{Sadiq Jaffer}
% \orcid{}
\affiliation{
	\institution{University of Cambridge}
	\country{United Kingdom}
}
\email{sj514@cam.ac.uk}

\author{Thomas Leonard}
\affiliation{
	\institution{University of Cambridge}
	\country{United Kingdom}
}
% \orcid{}
% \email{talex5@gmail.com}

\author{Jon Ludlam}
% \orcid{}
\affiliation{
	\institution{Tarides}
	\country{United Kingdom}
}
% \email{jon@tarides.com}

\author{Anil Madhavapeddy}
\orcid{0000-0001-8954-2428}
\authornotemark[1]
\affiliation{
	\institution{University of Cambridge}
	\country{United Kingdom}
}
\email{anil.madhavapeddy@cam.ac.uk}

% The default list of authors is too long for headers
\renewcommand{\shortauthors}{Gibb et al.}

\begin{abstract}
	Package managers are everywhere, with seemingly every language and operating system implementing their own solution.
	The lack of interoperability between these systems means that multi-lingual projects are unable to express precise dependencies across language ecosystems, and external system and hardware dependencies are typically implicit and unversioned.
	We define HyperRes, a formal system for describing versioned dependency resolution using a hypergraph that is expressive enough to model many ecosystems and solve dependency constraints across them.
	We define translations from dozens of existing package managers to HyperRes and comprehensively demonstrate that dependency resolution can work across ecosystems that are currently distinct.
	This does not require users to shift their choice of package managers; instead, HyperRes allows for the translation of packaging metadata between ecosystems, and for solving to be precisely specialised to a particular deployment environment.
\end{abstract}

\maketitle

\section{Introduction}
\label{sec:intro}

Almost every programmer uses package managers to manage their codebases, but they remain mysteriously ad-hoc and non-interoperable.
Package management software originally began with operating system distributions in the 1990s, popularised by Debian, Red Hat and Slackware.
These system package managers are responsible for distributing, assembling and managing filesystems which boot a distribution, and they typically focus on binary distributions to their users~\cite{4019575}.
A few years on, they were joined by language-specific package managers~\cite{10.1145/3365137.3365402,10.1145/1297081.1297093} that sit alongside the system package managers and offer more flexible and domain-specific interfaces to individual language ecosystems.
% Ryan: TODO say something about early repositories
A language package manager typically allows for selecting from multiple versions of a particular library, and for easier source-based compilation of a development branch.

Today, there is an explosion of both system and language package managers, each with seemingly incompatible interfaces.
This leads to all sorts of problems when trying to use packages from different ecosystems together.
Consider, for example, a project which uses a combination of C, Rust and OCaml code for a static binary, with more dynamic Python bindings that use GPU drivers that depend on a particular kernel driver.
Getting these (sometimes interlocked) dependencies resolved and stable is a Herculean task that is difficult to automate and keep reproducible.
The complexity led to the rise of containerization technologies like Docker, which sidestep the problem by bundling all dependencies together, but ultimately result in rather brittle deployments which are difficult to manage across ecosystems.
In our earlier example, four separate package managers are required to deploy this program: opam, Cargo, pip and APT.
If the program is to be portable beyond Debian, then even more will be required (e.g.~APK for Alpine, or RPM for Red Hat Linux).
Such a project might also break if the kernel GPU driver is updated, or if a Rust library is updated to a version that is incompatible with the particular version of Python used in the machine learning codebases.
If a security hole is found in one component, the whole stack must be upgraded very carefully indeed.
If continuous integration is deployed against the project, then all the package managers must be carefully managed to keep the deployment reproducible.
This situation is unergonomic, error-prone and deeply unsatisfying.

Yet all these package managers share a common core: they all resolve dependencies from a package repository and eventually output a filesystem subtree.
In this paper, we present a formal system that teases apart the seemingly monolithic features of package managers across both the operating system and programming language ecosystems, which we call the `hypergraph dependency resolution' (HyperRes) formalism.
HyperRes can not only be used to model many real-world package managers, but also forms the basis for the new ability to unify resolution requests across package management ecosystems with a single formula.
HyperRes can even be used to bidirectionally translate requirements directly into the package manager native formats, thus simplifying day-to-day management of multi-language and OS-portable software projects.
In the longer term, the HyperRes unified approach to version specification is vital towards securing software supply chains that are increasingly targeted by malicious actors.

%As a result there is a large design space in which they sit depending on implementation decisions or requirements of their use-case, yet there is a core commonality between them all.
%Despite this, these package managers do not generally interoperate.
%For example, if a portion of a codebase is written in another language and depends on that other language's compiler, there is normally no way to express this to the primary language's package manager.
%Programming language package managers can check for the presence of executables or libraries on systems, but there is little to no versioning for these and no guarantees that they will be compatible.
%As well as dependencies on packages typically provided by a system package manager, this also applies to cross-language libraries.
%There is no way to depend on a library from another ecosystem aside from duplicating it in the primary ecosystem (if the package manager is language-agnostic) or invoking the other ecosystems package manager.
%Package managers often sandbox builds which precludes the possibility of another package manager having network access, so dependencies are vendored into the project’s source directly, which increases compile times and leads to potential duplication.

%Take the example of an OCaml program using bindings to a Rust library deployed on a Debian Linux operating system.
%Three separate package managers will be required to deploy this program, opam, Cargo, and APT; the independently selected versions will have to be manually selected to be compatible.

HyperRes aims to extract a common core expressive enough to model numerous package managers, allowing resolution across their ecosystems.
In our earlier example, we would construct a single formula that would resolve OCaml, Python and Rust packages together with the system dependencies already available with a given Linux distribution.
HyperRes lifts the insight from earlier work~\cite{tucker2007opium} that package managers can be modelled as a constraint satisfaction problem, but extends it to a hypergraph structure that is suitable for bidirectional translation of the resolution formulae between ecosystems.
Thus, we can now even mechanically take a complex (e.g.) OCaml/Rust/Python project and create a binary \verb|deb| package that is binary-compatible with a given version of Debian Linux, by finding the common subset of packages that are already pre-packaged on that distribution vs those that need to be compiled from sources due to being only present in a particular package management ecosystem.
Today, this is a laborious and error-prone task that needs to be repeated per-distribution and per-architecture.
% Ryan: TODO we don't actually do this bi-directional translation, but we can say that we enable it?

In the remainder of this paper, we will first sketch out HyperRes informally~(\S\ref{sec:illustr}).
We then classify the commonality between many package managers~(\S\ref{sec:ontology}), and define the formal system for reasoning about dependency resolution from a resolution hypergraph to a resolved graph~(\S\ref{sec:formalism}).
We show how this system is expressive enough to model many real-world package managers~(\S\ref{sec:modelling}) and how we can use it as the basis to resolve dependencies across package management ecosystems~(\S\ref{sec:ecosystem}).
Finally, we consider related work and directions for future work~(\S\ref{sec:related}) and conclude~(\S\ref{sec:conclusion}).

%A package manager is a tool that automates the process of installing software packages.
%They can include executable binaries, shared objects, libraries, data, and more.
% The ergonomics can differ vastly; Nix is great for reproducible software deployment but cumbersome for rapidly changing software development due to expensive rebuilds hindering rapid iteration, and language-specific package managers

\subsection{An Illustrated Resolution Hypergraph}
\label{sec:illustr}

\begin{figure}
	\centering
	\includegraphics[width=\textwidth]{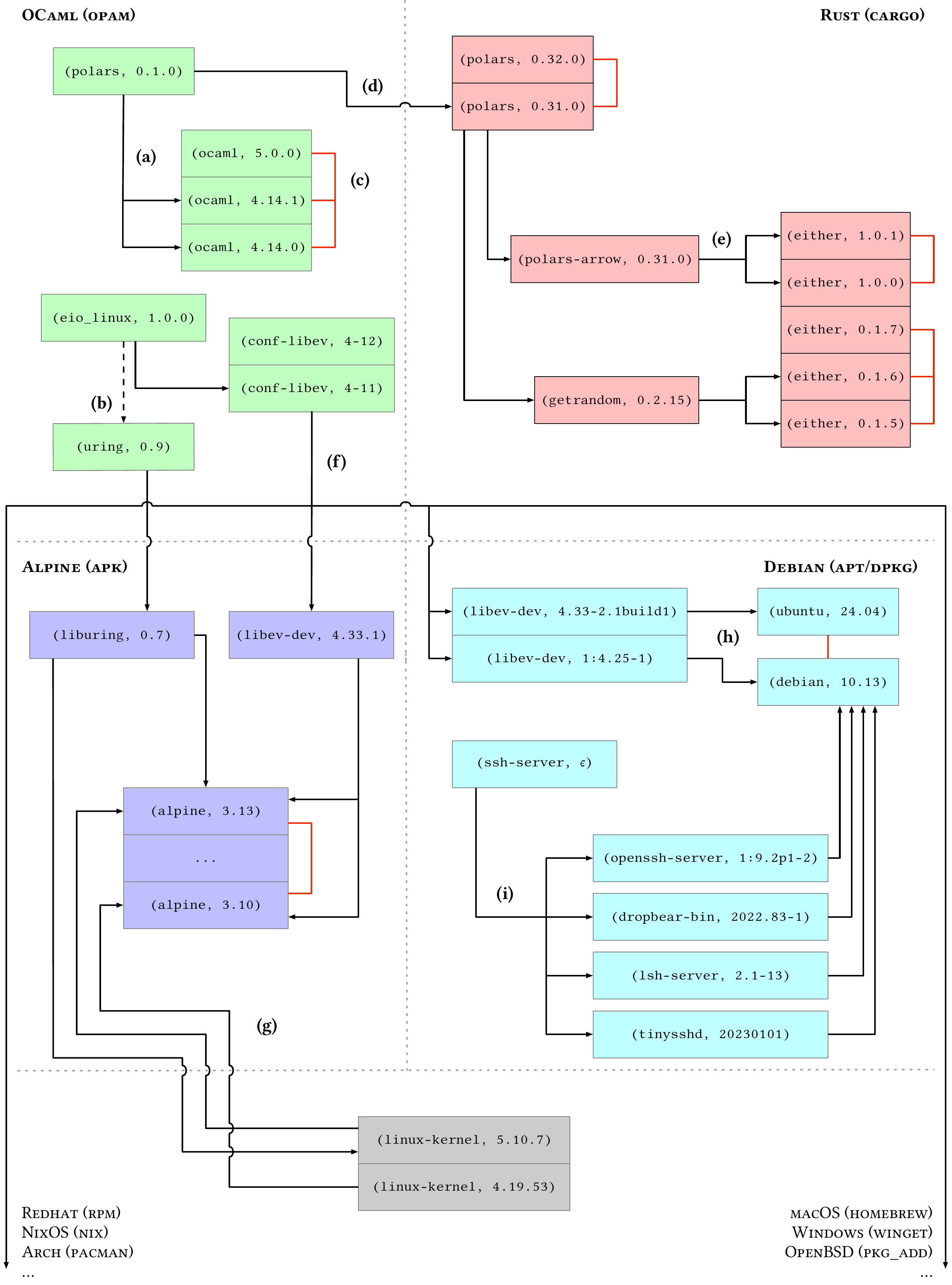}
	\captionsetup{justification=centering}
	\caption{An example cross-ecosystem hypergraph illustrating some of the challenges in package management.}
	\Description{TODO}
	\label{fig:packulus-miraculous}
\end{figure}
% Ryan: TODO tikz-ify this and add hyperrefs to the sections

We will first begin with a sketch of the HyperRes formalism~(\S\ref{sec:formalism}), with Figure~\ref{fig:packulus-miraculous} showing an example of a cross-ecosystem package hypergraph.
A hypergraph is a generalisation of a graph in which a `hyperedge' can join any number of vertices.
The diagram shows a hyperedge-labelled directed hypergraph representing a dependency resolution problem across ecosystems.
Vertices in this graph are packages, defined as a pair of package name and package version.
The packages included are real, but their relationships are chosen to illustrate interesting scenarios and to omit uninteresting incidental detail.

\subsubsection{Types of Dependencies}

Figure~\ref{fig:packulus-miraculous} shows three types of relationships between packages:
\paragraph{Strong dependencies.}
A hypergraph is a graph where `hyperedges' are between sets rather than individual vertices.
One such hyperedge represents `dependency' relationships where a package depends on a set of others, one of which can satisfy the dependency.
This is represented by a solid black line.
Note that we restrict the domain of directed hyperedges in our hypergraph to a size of 1 -- dependencies can only be \textit{from} one package.
For example, Figure~\ref{fig:packulus-miraculous} (a) shows \texttt{polars} depending on the \texttt{ocaml} compiler with a hyperedge from $(\{(\texttt{polars}, \texttt{0.1.0})\}, \{(\texttt{ocaml}, \texttt{4.14.1}), (\texttt{ocaml}, \texttt{4.14.0})\})$.
This means \texttt{polars} requires \textit{either} \texttt{ocaml} with version \texttt{4.14.0} or \texttt{4.14.1}.
In a package manager file format\footnote{In OCaml's case, this is the opam~\cite{opam} package manager; see Table~\ref{tbl:comparison} for details.} that is often expressed with a version formula such as \texttt{"ocaml" \{>= "4.14.0" \& < "5.0.0"\}}.
The fact that this relationship is a dependency is encoded in a labelling of the hyperedges.

\paragraph{Optional Dependencies.} Another such hyperedge represents `optional dependency' relationships, a set of packages which a package will use one of if present, but does not strongly require.
Figure~\ref{fig:packulus-miraculous}~(b) shows such a dependency from \texttt{(eio\_linux, 1.0.0)} to \texttt{(uring, 0.9)}, meaning that \texttt{eio\_linux} can use the Linux IO libraries if they are installed, but otherwise falls back to more portable \texttt{libev}-based ones.
Optional dependencies are commonly used to activate features across packages, such as enabling GPU support in a machine learning library if a GPU driver is present.

\paragraph{Conflicts.} The final package relationship type is a `conflict'.
In our running example, the OCaml toolchain only supports selecting a single version of a package, and so each \texttt{ocaml} package conflicts with every other \texttt{ocaml} package at Figure~\ref{fig:packulus-miraculous} (c), expressed as a (red) labelled hyperedge.
\begin{equation*}
\begin{gathered}
(\{(\texttt{ocaml}, \texttt{4.14.0})\}, \{(\texttt{ocaml}, \texttt{4.14.1}), (\texttt{ocaml}, \texttt{5.0.0})\})\\
(\{(\texttt{ocaml}, \texttt{4.14.1})\}, \{(\texttt{ocaml}, \texttt{4.14.0}), (\texttt{ocaml}, \texttt{5.0.0})\})\\
(\{(\texttt{ocaml}, \texttt{5.0.0})\}, \{(\texttt{ocaml}, \texttt{4.14.0}), (\texttt{ocaml}, \texttt{4.14.1})\})
\end{gathered}
\end{equation*}
For the simplicity of diagramming we represent this as an undirected `conflict set' where every package connected to the solid line conflicts with all the others.
Rust, in contrast, supports multiple versions of the same library and so does not need these conflicts for its packages.
Conflicts are usually introduced due to active software evolution such as incompatible APIs due to semantic versioning~\cite{6224274}, or due to some negative trust relationship between packages such as a security vulnerability or a bug in a particular version of a package~\cite{primiero2018negative}.

\subsubsection{Cross-Ecosystem Dependency Resolution}
\label{sec:illustration:cross-ecosystem}

The core dependency resolution problem~\cite{tucker2007opium} is based on the above three relationships.
It consists of mapping this hypergraph to a graph with concrete versions selected for each dependency, with optional dependencies selected if they are present in the graph due to another dependency, and with no conflicting packages.

We have described dependencies within the OCaml ecosystem so far, but Figure~\ref{fig:packulus-miraculous} (d) introduces a dependency across ecosystem boundaries from OCaml to the Rust \texttt{(polars, 0.31.0)} package.
Packages in different ecosystems are represented as vertices with a distinct namespace in the hypergraph, but that prefix is omitted in the diagram for clarity.
This is a novel feature of HyperRes, as existing package managers do not support cross-ecosystem dependencies.

\subsubsection{Multiple Versions of a Dependency}
\label{sec:illustration:concurrent-versions}

A transitive dependency of \texttt{(polars, 0.31.0)} is the \texttt{either} package, through \texttt{(polars-arrow, 0.31.0)} and \texttt{(getrandom, 0.2.15)}.
However, they are dependent on different versions of the \texttt{either} package, leading to a diamond dependency problem (Figure~\ref{fig:diamond-hypergraph}) where two versions of the same package are required to satisfy the dependency chain (Figure~\ref{fig:packulus-miraculous} (e)).
Some ecosystems forbid this behaviour (\S\ref{sec:table:concurrent-versions}) due to toolchain limitations or policy (due to the complexity of reasoning about the resulting graph).
Others, like the Rust ecosystem, support using multiple versions of the same library if they have different major versions, so these dependencies of \texttt{(polars, 0.31.0)} are satisfiable~\cite{cargoResolver}.

\subsubsection{Operating System Dependencies}
\label{sec:illustration:system-deps}

While cross-ecosystem dependency resolution is not supported by existing package managers, some such as opam have an `external dependency' mechanism for installing system dependencies by invoking the system package manager~\cite{opam}.
External dependencies are platform dependent, for example,
\begin{verbatim}
    ["libev-dev"] {os-distribution = "debian"}
    ["libev-dev"] {os-distribution = "alpine"}
\end{verbatim}
Figure~\ref{fig:packulus-miraculous} (f) shows this system dependency from \texttt{eio\_linux} on \texttt{libev-dev} from either Debian or Alpine Linux, depending on which operating system the package is being installed on.
For clarity, we omit the many possible, alternative systems that someone may be using.
These system dependencies are unversioned in existing package managers, and a frequent source of user-facing problems if they update their system package manager without updating the corresponding packages in the language package manager.
However, we can express these in HyperRes via cross-ecosystem dependencies~(\S\ref{sec:illustration:cross-ecosystem}).

% Ryan: TODO the kernel is a runtime dependency, which could perhaps be distinguished
At the lower end of the userspace software stack, it is also common to have packages that are tied to a particular OS kernel version or other hardware availability.
The \texttt{liburing} example in Figure~\ref{fig:packulus-miraculous} (g) shows how it depends not only on the Alpine OS version, but also the Linux kernel version.
Another notable example of this class of problem is with handling the injection of GPU drivers into the software dependency stack, which is a common requirement for machine learning software running on the cloud~\cite{10.1145/3530019.3530039}.

\subsubsection{Repository Selection}
\label{sec:illustration:repository-selection}

Debian Linux has package repositories associated with different releases of the operating system.
The APT package manager uses the repository associated with the currently installed Debian version, each of which contain distinctly versioned binary packages.
When upgrading the operating system, the package manager will select the repository associated with the new version and upgrade all packages to the new versions available in that repository.

HyperRes can identify which Debian release version satisfies a given package dependency query by modelling Debian and all of its releases as packages in the hypergraph.
Figure~\ref{fig:packulus-miraculous} (h) shows \texttt{(libev-dev,1:4.33-2.1build1)} depending on Ubuntu 24.04 Noble (a variant of Debian) and \texttt{(libev-dev,1:4.25-1)} depending on Debian 10.13 (Buster).

\subsubsection{Dependencies on Different Package Names}
\label{sec:illustration:different-names}

Except for our cross-ecosystem dependency in the Debian and Alpine namespaces~(\S\ref{sec:illustration:system-deps}), we have so far only expressed dependencies on packages of the same name.
However, several package managers (such as Debian) have a mechanism for expressing dependencies on virtual packages that can be satisfied by different implementation packages.
Figure~\ref{fig:packulus-miraculous} (i) shows one such package --- \texttt{ssh-server} --- that is provided by either \texttt{openssh-server}, \texttt{dropbear-bin}, \texttt{lsh-server}, or \texttt{tinysshd}.
These are represented naturally in the hypergraph a hyperedge to these various virtual packages.

Beyond this illustrative example, there are numerous other features of package managers that can be modelled in HyperRes, such as CPU architecture selection, operating system variants, feature flags for packages, optimisation flags, kernel version probing, and more.
These parameters, along with the hypergraph structured described earlier, all form the set of inputs to a resolver that can output a single, consistent, and reproducible set of packages that can be deployed to a machine.
 We will next dive into how we can break down real-world package managers into a more digestible form for conversion to and from the HyperRes formalism.

% something about traditional software deployment you deploy on the machine you have but in the age of VMs and cloud computing you can resolve to provision the exact configuration you need

\section{Finding the Common Pieces across Package Managers}
\label{sec:ontology}

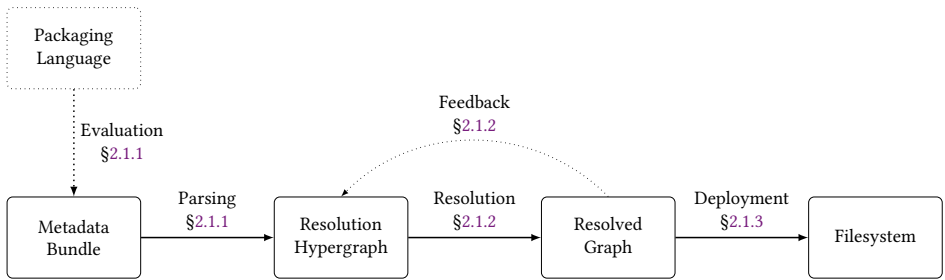
\begin{figure}
	\centering
	\resizebox{0.9\linewidth}{!}{\begin{tikzpicture} [
  node distance=5cm and 5cm,
  on grid,
  auto,
  >=latex,
  arrow/.style={draw, -latex, thick}
]

\begin{scope}[
  every node/.append style={
    rectangle,
    rounded corners=3pt,
    draw,
    align=center,
    minimum width=2.5cm,
    minimum height=1.5cm
  }
]
  \node[dotted]                (language)   {Packaging\\Language};
  \node[below=3.5 of language] (bundle)     {Metadata\\Bundle};
  \node[right=of bundle]       (hypergraph) {Resolution\\Hypergraph};
  \node[right=of hypergraph]   (graph)      {Resolved\\Graph};
  \node[right=of graph]        (filesystem) {Filesystem};
\end{scope}

\draw[dotted, thick, ->] (language)   -- (bundle)     node[align=center,midway,right] {Evaluation\\\S\ref{sec:pipeline:parsing}};
\draw[thick, ->]         (bundle)     -- (hypergraph) node[align=center,midway,above] {Parsing\\\S\ref{sec:pipeline:parsing}};
\draw[thick, ->]         (hypergraph) -- (graph)      node[align=center,midway,above] {Resolution\\\S\ref{sec:pipeline:resolution}};
\draw[thick, ->]         (graph)      -- (filesystem) node[align=center,midway,above] {Deployment\\\S\ref{sec:pipeline:deployment}};
\draw[dotted, ->, bend right=45] (graph.north) to     node[align=center,midway,above] {Feedback\\\S\ref{sec:pipeline:resolution}} (hypergraph.north);

\end{tikzpicture}}
	\captionsetup{justification=centering}
	\caption{The package management pipeline showing mappings between data structures,\\ with dotted lines denoting optional components.}
	\Description{
		A package management pipeline.
		The data structures are Packaging Language, Metadata Bundle, Resolution Hypergraph, Resolved Graph, and Filesystem.
		Evaluation maps from Packaging Language to Metadata Bundle, Parsing maps from Metadata Bundle to Resolution Hypergraph, Resolution.
	}
	\label{fig:pipeline}
\end{figure}

There are a lot of package managers in use in the wild, each with their own unique features and quirks, and the past two decades have seen a big surge in language-specific ones that sit alongside the more traditional system package managers.
But across them all, there are commonalities that can be used to build a reusable semantic core, leaving a smaller set of differences to be modelled as special cases.
We will now explore the design space of package managers by considering a representative selection of 36 out of a set of nearly 100 which we initially surveyed.\footnote{The keen reader may wish to skip ahead to Table~\ref{tbl:comparison} to see the full set of package managers we model using HyperRes.}
% RYAN: TODO I don't quite manage to model all the package manger's resolution steps in HyperRes (though it would make for good appendix material) so I might rein this claim in
Before we dive into the details, we will first define a common vocabulary for the typical stages in a package management pipeline as mappings between the data structures in Figure~\ref{fig:pipeline}.

\subsection{The Package Management Pipeline}

% Ryan: TODO collapse packaging language and metadata bundle into one, maybe Package Descriptions?

\subsubsection{Parsing}
\label{sec:pipeline:parsing}
Each package manager supports one or more repository formats, which are used to store the collection of available packages and consistency rules across them.
The `metadata bundle' is the collection of metadata associated with a given package that maps it into the package manager's repository format.
This is typically held outside the source code, since different package managers have different metadata formats and can update those independently of the releases of the downstream software.
For example, Debian's APT has a \texttt{control} archive in the \texttt{.deb} file format, OCaml's opam has \texttt{opam} files, Rust's Cargo has \texttt{Cargo.toml} files, and Nix~\cite{dolstraNixSafePolicyFree2004,10.1145/1411204.1411255} has store derivation \texttt{.drv} JSON files.
% Ryan: TODO say how metadata can be static or evaluated
% it's really the dependency information and build steps

This bundle may be contained in metadata files checked into version control for individual projects, and/or indexed in central repositories.
For instance, APT has a control file in \texttt{.deb} files along with an archive of the binary package, but also has a \texttt{Package} archive containing the same data for every package in a repository.
OCaml projects often have a \texttt{.opam} file checked into their version control repositories, and an opam repository contains the associated opam files for many projects, with a central \texttt{ocaml/opam-repository} representing the primary repository for that ecosystem.

Some package managers express their packages as embedded DSLs~\cite{10.1145/242224.242477} in a host language (such as Homebrew with Ruby).
%and Nix with the Nix DSL.
An `evaluation' step maps this packaging eDSL to the corresponding metadata bundle for that package (such as a JSON format in Homebrew).
% and \texttt{.drv} files in Nix.
Other package managers have a tighter integration between their packaging language and the bundle format, and define a concrete DSL~\cite{10.1145/6424.315691} exposed to users.
Cabal and opam, for example, each have their own (different) DSLs to describe packages, supporting features like variable assignment and boolean algebra which also includes references to evaluation steps such as build scripts.
% Ryan: TODO this section is a bit confused, due to the diversity between ecosystems. Just focus on the basics of dependency information.

The set of metadata bundles per ecosystem are then parsed into our resolution hypergraph.

\subsubsection{Resolution}
\label{sec:pipeline:resolution}
The resolution hypergraph is a representation of the dependency resolution problem: given a set of version constraints, how do we find the freshest set of packages that satisfy them?
The `resolution' step maps the hypergraph to a simpler `resolved' graph with concrete package versions selected.
This problem is NP-hard, as noted in earlier work~\cite{tucker2007opium,10.1145/2000229.2000255,10.1145/1858996.1859087}, but nowadays straightforward to solve using modern SAT solvers -- if the right cost functions are applied to the resolution process.
% RYAN: TODO add ref to the talex5 cost function magic section

% RYAN: section 3.2 proves this

Some package managers also take the existing installed state of the system into account when performing dependency resolution, denoted by the `feedback' loop from the resolved graph to hypergraph.
For example, Debian's supports an `apt-get upgrade' operation that modifies the version of a select set of packages to their latest versions, resolving any conflicts along the way.

Slackware is a notable exception to the norm in package management dependency resolution; official mechanisms do not provide any support for dependency tracking, instead relying on the user to manually install dependencies of a package.
This reflects that distribution's philosophy of simplicity and user control over a more rigorous dependency resolution system as found in (e.g.) Debian or Red Hat.

\subsubsection{Deployment}
\label{sec:pipeline:deployment}
The final mapping is from the resolved package graph to a filesystem subtree that can be deployed.
In some specialised cases (such as embedded systems such as Busybox or Yocto), the deployment step may be to a firmware image rather than a filesystem, but the principle remains the same.
% RYAN: are there any package managers that don't use the filesystem? And install things only in-memory?
% ANIL: clarified above.
The specifics of the deployment steps varies on the nature of the package manager.
Source-based language package managers for dynamic languages such as JavaScript might simply place source code into a local directory, as npm does.
System package managers that handle binary packages might unpack archives into the file system hierarchy.
Other package managers might build the software according to instructions in the package metadata, as happens with cargo, opam or cabal.

Package managers might also make network requests to download source or binary code during deployment if a local repository is not present (e.g.\ opam), or have all the source code provided directly within the metadata bundle format (e.g.\ APT).

The specifics of how package managers manage resolved graphs varies, depending on how the clients handle state management.
OCaml's opam uses global or project-local `switches', allowing multiple resolved graphs to live-side-by side~\cite{opam}.
JavaScript's npm puts sources into a project-local \verb|node_modules/| directory.
APT uses dpkg to unpack prebuilt packages into the Linux Filesystem Hierarchy Standard (FHS).
Nix places built packages at a path containing a cryptographic hash of their store derivation in the `Nix store' (usually found in \verb|/nix|).
% RYAN: we could link to section 4.4 here, but we're mostly avoiding forward linking in favour of back linking unless it's required for clarification (e.g. saying we'll explain the formalism in section~\ref{sec:formalism}.
Meanwhile, Haiku OS deploys packages by mounting them in a union filesystem, allowing for layering and sharing of multiple packages.
% RYAN: we could talk a bit more about what properties to these support? E.g. atomic upgrades/downgrades, reproducible deployments

% THOMAS: what about "Evaluation" and "Feedback" - they appear on the graph but are not explained
% RYAN: I've added links from the graph to the parsing and resolution sections respectively

\newcommand\chk{\color{black}{\ding{51}}}
\newcommand\crs{\color{red}{\ding{55}}}
\newcommand\naa{-}

% NB if using vimtex you might want
% `let vimtex_indent_enabled=0`

\afterpage{

\let\oldfootnotesize\footnotesize
\renewcommand{\footnotesize}{\fontsize{6pt}{0pt}\selectfont}

% \newgeometry{
%     top=58pt, bottom=44pt, inner=46pt, outer=46pt
% }
\thispagestyle{empty}
\begin{landscape}
\begin{table}
\vspace{-2em}
\fontsize{6pt}{8pt}\selectfont
\centerline{
\begin{tabular}{llll|lllll|lllll}
\multicolumn{4}{c|}{Description} & \multicolumn{5}{c|}{Functionality} & \multicolumn{5}{c}{Pipeline}\\
\bf \makecell{Package\\Manager\\Name}
& \bf \makecell{Number of\\Packages\tablefootnote{Package name and version pairs in the primary repository rounded to the nearest significant figure.}\\{\em (magnitude)}}
& \bf \makecell{Open Source\\Ecosystem\\\S\ref{sec:table:ecosystem}}
& \bf \makecell{Release\\Cycle\\\S\ref{sec:table:repository}}
& \bf \makecell{Binary\\Distribution\\\S\ref{sec:table:binary}}
& \bf \makecell{Dependency\\Formula\\\S\ref{sec:table:dependency-formula}}
& \bf \makecell{Concurrent\\Versions\\\S\ref{sec:table:concurrent-versions}}
& \bf \makecell{Toolchain\\Integration\tablefootnote{Whether a tool is a \bcircle{P}ackage manager, \bcircle{B}uild system, and/or a \bcircle{C}ompiler.}\\\S\ref{sec:table:toolchain}}
& \bf \makecell{Sandboxed\\Builds\\\S\ref{sec:table:sandboxing}}
& \bf \makecell{Packaging\\Language\\\S\ref{sec:pipeline:parsing}}
& \bf \makecell{Metadata\\Bundle\\\S\ref{sec:pipeline:parsing}}
% Ryan: TODO combine packaging language and metadata bundle
& \bf \makecell{Resolution\\Hypergraph\\\S\ref{sec:pipeline:resolution}}
& \bf \makecell{Resolved\\Graph\\\S\ref{sec:pipeline:resolution}}
% Ryan: TODO this distinction doesn't make sense for most package managers -- rename to `resolution'
& \bf \makecell{Filesystem\\State\\\S\ref{sec:pipeline:deployment}}
% Ryan: TODO rename as `deployment', basically focusing on the mappings from figure 2 rather than the data structures
\\
\hline
%name                               & no.            & ecosystem          & release   & bin               & vers & con  & tool             & sand & language         & bundle                          & hyper                 & graph                 & filesystem
Maven                               & $2\times10^7$  & Java               & Rolling   & Binary            & \chk & \chk & \bcircle{P} \bcircle{B}
	                                                                                                                                       & \crs & \crs             & \texttt{pom.xml}                & \texttt{maven}        & \texttt{maven}        & Local Repository     \\
% https://mvnrepository.com/repos
pip                                 & $6\times10^6$  & Python             & Rolling   & Binary            & \chk & \crs & \bcircle{P}      & \crs & \crs             & \texttt{pyproject.toml}\tablefootnote{pip previously used \texttt{requirements.txt}.}
	                                                                                                                                                                                                   & \texttt{pip}          & \texttt{wheel}        & virtual environments \\ % https://pypi.org/
Poetry                              & $6\times10^6$  & Python             & Rolling   & Source            & \chk & \crs & \bcircle{P}      & \crs & \crs             & \texttt{pyproject.toml}         & \texttt{poetry}       & \texttt{poetry.lock}  & Virtual environment  \\
npm                                 & $3\times10^6$  & Javascript         & Rolling   & Source            & \chk & \chk & \bcircle{P}      & \crs & \crs             & \texttt{package.json}           & \texttt{npm}          & \texttt{package.lock} & \texttt{node\_modules}\\
Yarn                                & $3\times10^6$  & Javascript         & Rolling   & Source            & \chk & \chk & \bcircle{P}      & \crs & \crs             & \texttt{package.json}           & \texttt{yarn}         & \texttt{yarn.lock}    & \texttt{node\_modules}\\
Cargo~\cite{cargo}                  & $1\times10^6$  & Rust               & Rolling   & Source            & \chk & \chk & \bcircle{P} \bcircle{B}\tablefootnote{Cargo is a package manager and build system.}
																																		   & \crs & \crs             & \texttt{Cargo.toml}             & \texttt{cargo}        & \texttt{cargo}        & Cargo cache          \\
Spack~\cite{spack}                  & $4\times10^5$  & HPC                & Versioned & Evaluated         & \chk & \chk & \bcircle{P}      & \chk & Python           & \texttt{package.py}             & \texttt{clingo}       & \texttt{spack}        & \texttt{SPACK\_ROOT} \\
Conda                               & $3\times10^5$  & Python             & Rolling   & Source            & \chk & \crs & \bcircle{P}      & \crs & \crs             & \texttt{meta.yml}               & \texttt{conda}        & \texttt{conda}        & Conda environment    \\ % https://conda-forge.org/ package names
% spack $ cat **/package.py | grep "version(\"" | wc -l
CPAN                                & $2\times10^5$  & Perl               & Rolling   & Source            & \chk & \crs & \bcircle{P}      & \crs & \crs             & \texttt{cpanfile}               & \texttt{cpan}         & \texttt{cpan}         & \texttt{~/.cpan}     \\
Chocolatey                          & $2\times10^5$  & Windows            & Rolling   & Binary            & \chk & \crs & \bcircle{P}      & \crs & \crs             & \texttt{.nuspec} (\texttt{XML}) & \texttt{.nuspec} \tablefootnote{Installation sources (\texttt{.msi},\texttt{.msix},\texttt{.appx},\texttt{.exe}) may manage their own dependencies.}
																																																							   & \texttt{choco}       & \texttt{Program Files} \\
Gem                                 & $2\times10^5$  & Ruby               & Rolling   & Source            & \chk & \crs & \bcircle{P}      & \crs & Ruby             & \texttt{Gemfile.lock}           & \texttt{bundle}       & \texttt{bundle}       & A RubyGems directory \\
Cabal~\cite{haskell}                & $1\times10^5$  & Haskell            & Rolling   & Source            & \chk & \crs & \bcircle{P} \bcircle{B}\tablefootnote{Cabal is a package manager and build system.}
																																		   & \chk & \texttt{.cabal}  & \texttt{.cabal}                 & \texttt{cabal}        & \texttt{ghc-pkg}      & Cabal cache          \\
Nix~\cite{dolstraNixSafePolicyFree2004}                                 & $9\times10^4$  & Various            & Versioned\tablefootnote{Nix channels act as versioned repositories.}
                                                                                      & Evaluated         & \crs & \chk & \bcircle{P}      & \chk & Nix expressions  & store derivations               & Nix realise           & Nix realise           & Nix store            \\
RPM                                 & $8\times10^4$  & Fedora             & Versioned & Binary            & \crs & \crs & \bcircle{P}      & \crs & \crs             & \texttt{.rpm}                   & \texttt{dnf}          & \texttt{rpm}          & FHS                  \\
APT                                 & $6\times10^4$  & Debian Linux       & Versioned & Binary            & \chk\tablefootnote{Debian often doesn't specify dependency versions and relies on a repository keeping a compatible package set.}
																											     & \crs & \bcircle{P}      & \crs & \crs             & \texttt{.deb}                   & \texttt{apt}          & \texttt{dpkg}         & FHS                  \\
RPM                                 & $5\times10^4$  & OpenSUSE           & Versioned & Bundles           & \crs & \crs & \bcircle{P}      & \crs & \crs             & \texttt{.rpm}                   & \texttt{zypper} using \texttt{libsolv} & \texttt{rpm}          & FHS                  \\
winget                              & $5\times10^4$  & Windows            & Rolling   & Binary            & \crs & \crs & \bcircle{P}      & \crs & \crs             & YAML Manifest                   & \texttt{.installer.yaml} \tablefootnote{Installation sources (\texttt{.msi},\texttt{.msix},\texttt{.appx},\texttt{.exe}) may manage their own dependencies.}
																											                                                                                                                   & \texttt{winget-cli}   & \texttt{Program Files} \\
\texttt{pkg}                        & $4\times10^4$  & FreeBSD            & Versioned & Binary            & \crs & \crs & \bcircle{P}      & \crs & \crs             & \texttt{.pkg}\tablefootnote{\texttt{.tgz} file format with JSON manifest.}
                                                                                                                                                                                                       & \texttt{pkg}          & \texttt{pkg}          & FHS                  \\
opam~\cite{opam}                    & $3\times10^4$  & OCaml              & Rolling   & Source            & \chk & \crs & \bcircle{P}\tablefootnote{Opam supports language-agnostic build scripts.}
																																		   & \chk & opam file        & opam file                       & \texttt{opam}         & \texttt{opam}         & Opam switch          \\
Guix~\cite{courtes2013functional}   & $3\times10^4$  & Various            & Rolling   & Evaluated         & \crs & \chk & \bcircle{P}      & \chk & Scheme           & store derivations               & Guix build            & Guix build            & Guix store           \\
Zero Install                        & $3\times10^4$  & Cross-platform     & Rolling   & Binary            & \chk & \chk & \bcircle{P}      & \crs & \crs             & \texttt{feed.xml}               & \texttt{0install select}
									                                                                                                                                                                                           & \texttt{selections.xml} & 0install cache       \\ % http://roscidus.com/0mirror/users/88C8A1F375928691D7365C0259AA3927C24E4E1E/user.html
pacman                              & $2\times10^4$  & Arch Linux         & Rolling   & Binary            & \crs & \crs & \bcircle{P}      & \crs & \crs             & \texttt{PKGBUILD}               & \texttt{pacman}       & \texttt{pacman}       & FHS                  \\
APK                                 & $2\times10^4$  & Alpine Linux       & Versioned & Binary            & \chk & \crs & \bcircle{P}      & \crs & \crs             & \texttt{.apk}                   & \texttt{apk}          & \texttt{apk}          & FHS                  \\
Portage                             & $2\times10^4$  & Gentoo             & Rolling   & Source            & \chk & \chk\tablefootnote{Limited support with `Slotting'.}
																														& \bcircle{P}      & \chk & ebuild scripts   & ebuild scripts                  & \texttt{emerge}       & \texttt{emerge}       & FHS                  \\ % https://packages.gentoo.org/  19054 package names
CRAN                                & $2\times10^4$  & R                  & Versioned & Source            & \crs & \crs & \bcircle{P}      & \crs & \crs             & \texttt{DESCRIPTION}            & \texttt{packages}\tablefootnote{CRAN packages are installed from invoking an R function in the source of a package.}
																																																							   & \texttt{packages}     & Local directory      \\ % packages names https://cran.r-project.org/
TLmgr                               & $8\times10^3$  & TeX Live           & Versioned & Source            & \crs & \crs & \bcircle{P}      & \crs & \crs             & \texttt{tlpkg} directory        & \texttt{tlmgr}        & \texttt{tlmgr}        & texmf trees          \\ % pkg names possibly
Homebrew                            & $7\times10^3$  & macOS              & Rolling   & Binary            & \crs & \chk \tablefootnote{By manual name mangling, and only done for $~1\%$ of packages.}
                                                                                                                        & \bcircle{P}      & \crs & Ruby             & Formula                         & \texttt{brew}         & \texttt{brew}         & Cellar               \\
pkgman                              & $4\times10^3$  & Haiku OS           & Versioned & Binary            & \chk & \crs & \bcircle{P}      & \crs & \crs             & \texttt{.PackageInfo}           & \texttt{pkgman}       & \texttt{pkgman}       & Union FS             \\ % package names
Stack                               & $3\times10^3$  & Haskell            & Versioned & Source            & \chk & \crs & \bcircle{P} \bcircle{B}
																																		   & \chk & \crs             & \texttt{.cabal}                 & \texttt{stack}        & \texttt{ghc-pkg}      & Stack cache          \\
pkgtools~\cite{slackware}           & $2\times10^3$  & Slackware          & Versioned & Binary            & \crs & \crs & \bcircle{P}      & \crs & \crs             & \texttt{.tgz}                   & \crs                  & \crs\tablefootnote{Slackware doesn't track dependencies.}
                                                                                                                                                                                                                                                       & FHS                  \\
RPM                                 & $2\times10^3$~\tablefootnote{Without subscription.}
                                                     & Red Hat Linux      & Versioned & Binary            & \crs & \crs & \bcircle{P}      & \crs & \crs             & \texttt{.rpm}                   & \texttt{dnf}          & \texttt{rpm}          & FHS                  \\
\texttt{pkg\_add}                   & $1\times10^3$  & OpenBSD            & Versioned & Binary            & \crs & \crs & \bcircle{P}      & \crs & \crs             & \texttt{.tgz}                   & \texttt{pkg\_add}     & \texttt{pkg\_add}     & FHS                  \\
PEAR                                & $6\times10^2$~\tablefootnote{We only count the number of unique package names in PEAR as versioning information was not accessible.}
													 & PHP                & Versioned & Source            & \chk & \crs & \bcircle{P}      & \crs & \crs             & \texttt{package.xml}            & \texttt{pear}         & \texttt{pear}         & PEAR directory       \\
Go modules~\cite{goModules}         & n/a\tablefootnote{Go doesn't have a central repository.}
	                                                 & Go                 & n/a\tablefootnote{Go doesn't use repositories by default, but can be configured to do so.}
                                                                                      & Source            & \crs\tablefootnote{Go uses Minimal Version Selection~\cite{coxGoMVS2018}.}
                                                                                                                 & \chk \tablefootnote{The Go tool is a package manager, build system, and compiler.}
                                                                                                                        & \bcircle{P} \bcircle{B} \bcircle{C}
																														                   & \crs & \crs             & \texttt{go.mod}                 & \texttt{go}           & \texttt{go.sum}       & Go Module Cache      \\
Bazel                               & n/a\tablefootnote{Bazel doesn't have a software repository, instead declaring `external dependencies'.}
                                                     & Multi-language     & n/a       & Source            & \chk & \crs & \bcircle{P} \bcircle{B}\tablefootnote{Bazel is a build system with package management functionality.}
																																		   & \chk & \crs             & \texttt{WORKSPACE} file         & \texttt{bazel}        & \texttt{bazel}        & Bazel directory      \\
%?                                  & ?              & ?                  & ?         & ?                 & ?    & ?    & ?                & ?     & ?               & ?                               & ?                     &?                      & ?                    \\
% bazel
% buck
% pants
% flatpak
% snap
% devcontainers
% AppImage
% XARs
% Bedrock Linux
% vim plugins
% winget
% chocolate
% store
% Docker
\end{tabular}
}
\captionsetup{justification=centering}
\caption{A comparison of representative package managers.}
\label{tbl:comparison}
\vspace{-1in}
\end{table}
\end{landscape}

\renewcommand{\footnotesize}{\oldfootnotesize}

% \restoregeometry

}

% RYAN: TODO make section titles match table headers

\subsection{A Categorisation of Package Managers}
\label{sec:table}

Table~\ref{tbl:comparison} fits the numerous package managers we surveyed into the pipeline in Figure~\ref{fig:pipeline}.
The table is a representative sample of interesting properties across the much larger set that we surveyed.
For the purposes of brevity, we preferred open source ecosystems and did not include `app store' managers like Google Play or Apple's App Store, nor did we include package managers that are not primarily used for software deployment, such as \TeX's CTAN.

%We also characterise package numbers with a number of other axes.

\subsubsection{Ecosystems}
\label{sec:table:ecosystem}
The first section of Table~\ref{tbl:comparison} describes the particular use case a package manager was created for.
As noted, package managers generally fall into two categories: language package managers and system package managers.
% Language package managers are built to manage dependencies for a specific programming language, and system package managers are used to manage packages in an Operating System.
The intended use for language package managers is in distributing libraries and tools to developers in a specific language, and system package managers in administrating a full operating system.
The lines between these often blur; systems package managers do distribute language libraries, and language package managers can facilitate setting up a development environment with system dependencies.
The system package managers tend to lag behind in the latest versions of libraries as they are more conservative in their updates, while language package managers typically offer less consistent integration across each other.

% how does dependency resolution work in system package managers compared to language package managers?
% - dependency resolution still happens, it is just that multiple versions of a package are not distributed in one repository (where a repository is a `version' of system)
% - hackage is an example of bringing this philosophy to a language package manager

\subsubsection{Repositories}
\label{sec:table:repository}
Packages managers almost always group packages together in a centralised repository, allowing for discoverability and upgrades.
% language (e.g. Nix) or bundle
Some package managers keep one main repository where changes are made continuously, dubbed a `rolling release'.
Others periodically release a version of their repository, much as a new package version is released, representing a `versioned release' of their operating system distribution.
Security and other updates are usually backported to previous release of their repository for some time, creating `minor' releases which should be safe for every user to upgrade to without risking software incompatibilities.
Multiple repositories can be also often be combined with, for example,~non-free packages, or unofficial user-contributed packages.
% Nix channels are versions of the Nixpkgs repositories
% Nix flakes are other repositories
A notable exception to this structure that does not require a package repository is Go, instead pointing to packages via a URL.

\subsubsection{Bundling Mechanism}
\label{sec:table:binary}
% Ryan: TODO I don't like this terminology... let's go with `deployment type', one of source, binary, or build
All package managers wrap an existing set of source code with some metadata to eventually install some entries into a filesystem, but the specifics can vary.

\paragraph{Source bundles.}
Some package managers only deal with deploying the sources for packages which are then built as part of a project, or interpreted.
For example, the Go module system assembles all the sources before invoking its integrated build system.
%~(\S\ref{sec:table:toolchain}).
Others, such as npm, assemble the source packages (e.g. of JavaScript, an interpreted language) along with C bindings and other conventionally compiled files, and support builds via pre- or post-publishing scripts that can, for example, transpile from TypeScript to ES5~\cite{npmScripts} or invoke a C compiler.
Other approaches build projects using a shell script from the package metadata as part of deploying them~(\S\ref{sec:pipeline:deployment}), like opam, Portage, and Nix.
All of these are denoted as `source' in Table~\ref{tbl:comparison}.

\paragraph{Binary bundles.}
Some package managers --- typically the operating system oriented ones --- specifically have prebuilt binary packages as part of the bundle format.
%This fits into the package management pipeline as a pre-processing step.
%Dynamic dependencies?
APT and RPM are two such examples; they begin their package lifecycle via source packages that are compiled to create multiple binary packages which are then installable by the end-user.
These can be viewed as a staged version of source bundles, but with a concrete binary packaging format that is optimised for deployment for the resulting binaries that uses a simpler version specification than the source packages (\S\ref{sec:pipeline:resolution}).
These are all denoted `binary' in Table~\ref{tbl:comparison}.

\paragraph{Evaluated bundles.}
Nix or Guix~\cite{courtes2013functional} are a combination of the above two schemes: they build projects in a sandboxed environment via a computed package specification DSL, with precisely specified inputs that track their dependency chains as the DSL is evaluated.
This results in a single binary deployment of the collection of selected packages as a transparent optimisation of source deployment, but potentially requires much more recomputation by the end-user than a binary system.
These are denoted as `evaluated' in Table~\ref{tbl:comparison}.

\subsubsection{Dependency Formula}
\label{sec:table:dependency-formula}
Version formulae are used by a package to express a set of possible packages that can satisfy a dependency.
For example, a package depending on $1 \leq \texttt{a} \leq 3$ might depend on a set containing \texttt{a.1}, \texttt{a.2}, and \texttt{a.3}.
Some package managers have limited dependency formulae consisting of a single range, others have arbitrarily complex boolean algebra, and others still support only specifying a \textit{single} package that can satisfy a dependency.
We differentiate between those that support expressing a dependency on a set of possible packages versus those that can only depend on an exact package in the `dependency formula' field of the table.
% An example of the latter is Nix.

The `packaging language' and `metadata bundle' fields show the languages and formats described in Section~\ref{sec:pipeline:parsing}.
The `resolution hypergraph' and `resolved graph' fields show the tooling typically used to interact with these data structures.
For example, \texttt{apt} is used for dependency resolution and, with a topological sort over the resolved graph, the lower-level \texttt{dpkg} is used to deploy individual binary packages.
Finally, the `filesystem state' field shows how the packages are deployed to the filesystem.
% Ryan: TODO update this with new fields

\subsubsection{Concurrent Versions}
\label{sec:table:concurrent-versions}
%For example...
Some package managers allow only one version of a package name in a resolved graph due to deployment constraints such as requiring unique symbols when linking objects, or storing packages at a unique path in the filesystem.
For example, the OCaml toolchain does not support multiple versions of a package in the same switch.
As such, while opam supports multiple resolved graphs living side-by-side in separate switches~(\S\ref{sec:pipeline:deployment}), it doesn't support multiple versions in one dependency resolution.
System package managers often put a package at a particular path in the filesystem, so multiple versions of the same package would collide.
% Ryan: TODO mention Gentoo slots but also their limitations

Other package managers allow multiple versions of the same package name existing in a resolved graph, `concurrent versions', by supporting deploying multiple versions of the same package name.
% THOMAS: minor note, but opam could easily support installing multiple versions in the same switch. The issue is that ocamlfind and the compiler do not so opam restrict this.
% RYAN: good point, I've added a better explanation
Cargo uses name mangling in order to support linking multiple versions of a single library into the same binary, where typically only one version can be used at the same time due to duplicate symbols.
Note that Cargo only does this for packages with different major package versions as part of the semantic versioning scheme~\cite{cargoResolver, semver}.
Nix diverges from the Linux Filesystem Hierarchy Standard\cite{lsb} to packages at paths containing unique cryptographic hashes, allowing the installation of multiple versions of a package on the same system.

\subsubsection{Toolchain Integration}
\label{sec:table:toolchain}
Some package managers, especially language package managers, have some degree of integration with other parts of the toolchain; build systems and compilers.
Opam includes build scripts in package metadata which can invoke a build system or compiler, but is language-agnostic.
This is similar to how a build system's rules will include invocations to a compiler, such as a makefile target invoking \texttt{gcc}.
Other ecosystems have tighter integration; cargo is the Rust ecosystem's package manager \texttt{and} build system.
The Go tool is a package manager, build system, and compiler.
This enables functionality that might not be otherwise possible without a well-defined API between these parts of the toolchain.
For example, Cargo's concurrent version support is enabled by using knowledge of package management to provide unique symbols using package versions.

\subsubsection{Sandboxing}
\label{sec:table:sandboxing}
Sandboxing is employed by package managers to ensure reproducibility and isolation during the build process.
For example, opam and Nix sandbox builds when building packages for deployment~(\S\ref{sec:table:binary}).
The technologies used for sandboxing are often similar to those employed in containerisation and virtualisation.
For example, Linux namespaces allow for the creation of lightweight environments where builds have access only to explicitly defined resources, such as specific file systems, process trees, and network interfaces.
Sandboxing in package management is distinct from \textit{reproducible builds}, which aim for bit-for-bit identical binaries and require determinism in all parts of the toolchain~\cite{reproducibleBuilds}.

% TODO how can we build without invoking the package manager
%Bazel the build system does package management
% http://blog.ezyang.com/2015/12/the-convergence-of-compilers-build-systems-and-package-managers/

%Other software deployment tools have package-manager like functionality.
%Docker manages container images similar
%Flatpack and so on install applications sandboxed.

\section{A Formal System for Dependency Resolution}
\label{sec:formalism}

In this section, we define the HyperRes formalism that allows us to reason about dependency resolution and extract a common core to package management.
This addresses the resolution mapping~(\S\ref{sec:pipeline:resolution}) from Figure~\ref{fig:pipeline}.

\subsection{HyperRes}
\label{sec:hyperres}

\begin{wrapfigure}[30]{R}{0.25\textwidth}
	\captionsetup{justification=centering}
	\begin{minipage}{\linewidth}
		\centering
		\includegraphics[width=\textwidth]{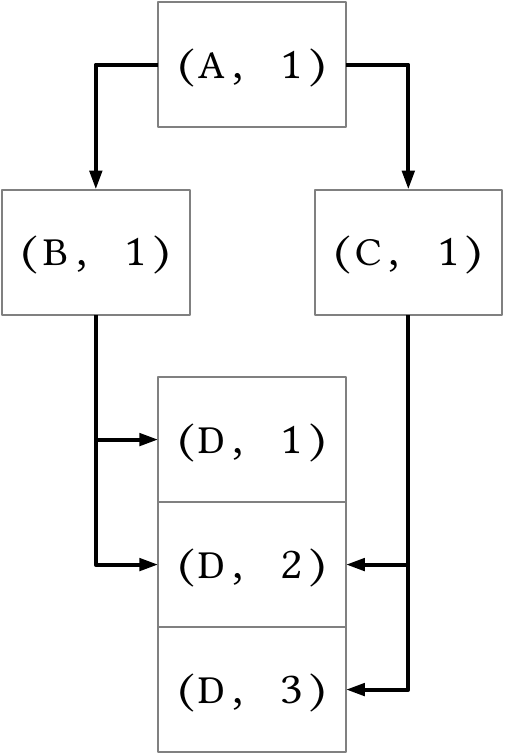}
		\subcaption{A resolution hypergraph $H$.}
		\Description{A resolution hypergraph $H$ where,
			$P=\{A1, B1, C1, D1, D2, D3\}$,
			$R=\{
				(\{A1\},\{B1\}),
				(\{A1\},\{C1\}),
				(\{B1\},\{D1,D2\}),
				(\{C1\},\{D2,D3\})
			\}$, and
			$\forall r \in R, L(r)=\delta$.
		}
		\label{fig:hypergraph}
		\vspace{0.2in}
		\includegraphics[width=\textwidth]{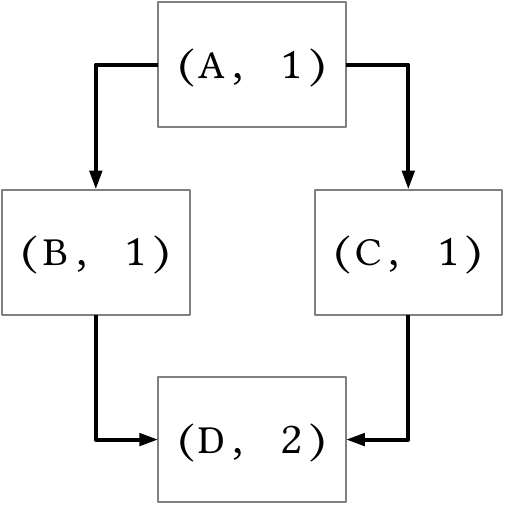}
		\subcaption{A resolved graph $G$.}
		\Description{A resolved graph $G$ where $q=A1$,
			$V(G)=\{A1, B1, C1, D2\}$, and
			$E(G)=\{
				(A1,B1),
				(A1,C1),
				(B1,D2),
				(C1,D2)
			\}$.
		}
		\label{fig:graph}
	\end{minipage}
	\caption{Example of a resolution from $H$ and $q=A1$ to $G$.}
	\label{fig:hypergraph-to-graph}
\end{wrapfigure}

The \textit{resolution hypergraph} is a hyperedge-labelled directed hypergraph, where a hyperedge represents a relationship from one package to a set of packages.
The type of this relationship is denoted by a hyperedge-label and is one of `dependency', `optional dependency', or `conflict'.
The resolution hypergraph is resolved to a subgraph which satisfies the dependency constraints in a process we refer to as dependency resolution.

% THOMAS: to discuss - I think we can simplify this a bit - here you are not stating that every package in $d$ has the same name (which is probably what you wanted, given how you use this after).
% RYAN: I actually did this intentionally. This means, for example, that we can express a constraint on one of a number of packages that provide a dependency in Debian. E.g. (openssh-server | dropbear-bin)
% THOMAS: But actually, I think this is better to not enforce it and also remove the 'exists' constraint. I would do something like this:
% - we want the dependency formulas to be as generic as possible
% - we want our model to manipulate them with a CNF form (you can transform it in polynomial space, - cite Karp)
% - we use an hypergraph to model the package relationships - node are packages, edges are derived from those CNF clauses
% RYAN: so I was thinking about a couple of scenarios here,
% (1) if we have openssh-server and dropbear-bin which are co-installable (don't conflict), without this constraint we could get both in $G$ where we only need one
% (2) if the case of cargo, if we depend on D1, D2, and D3, without the uniqueness constraint we could include all of these in $G$ where we only need one
To start, we define a set of package names $N$, and a set of versions $V_n$ for every package name $n$ in $N$.
The set of packages is often expressed with \textit{dependency formula}~(\S\ref{sec:table:dependency-formula}).
Though there are many versioning schemes out there, we do not attach any particular semantics or ordering to versions.
We define a set of all packages as name-version pairs,
\begin{wrapped}{align*}
P = \left\{ \left(n, v\right) \mid n \in N, v \in V_n \right\}
\end{wrapped}

A package $p \in P$ can express a dependency on a set of packages $d \subseteq P$, one of which can satisfy the dependency.
Package $p$ can have multiple dependencies, all of which must be satisfied.
Similarity, a package $p$ can express an optional dependency on a set of packages $o \subseteq P$, one of which can satisfy the dependency; and a conflict on a set of packages $c \subseteq P$ which $p$ cannot co-exist with.
We define functions $deps~:~P~\to~2^{2^P}$ mapping a package to the set of its dependencies, $opts~:~P~\to~2^{2^P}$ mapping a package to set of its optional dependencies, and $conflicts~:~P~\to~2^{2^P}$ mapping a package to the set of its conflicts.
We define a set of ordered pairs of relationships as,
\begin{wrapped}{align*}
	R = & \left\{ \left(\left\{p\right\}, d\right) \in 2^P \times 2^P \mid p \in P, d \in deps(p) \right\} \\
	\cup & \left\{ \left(\left\{p\right\}, o\right) \in 2^P \times 2^P \mid p \in P, o \in opts(p) \right\} \\
	\cup & \left\{ \left(\left\{p\right\}, c\right) \in 2^P \times 2^P \mid p \in P, o \in conflicts(p) \right\}
\end{wrapped}

We define the resolution hypergraph as a directed hypergraph $H$ with vertices $P$ and hyperedges $R$, where hyperedges from a package (the domain) to a set of packages (the codomain)~\cite{bergeHypergraphs}.
Note that we restrict the domain to a size of one -- we can only express a dependency \textit{from} one package.

We define labels for the hyperedges of our hypergraph as a mapping $L~:~2^P~\times~2^P~\to~T$ from hyperedges to a relationship $T=\{\delta, \sigma, \gamma\}$, where $\delta$~encodes a dependency, $\sigma$~an optional dependency, and $\gamma$~a conflict.
% i.e. we can only map to one
Note that this means we cannot depend, optionally depend, or conflict with the same package set.
\begin{align*}
	\forall p \in P, \forall d \in deps(p), L\left(\left(\left\{p\right\}, d\right)\right) = \delta \\
	\forall o \in opts(p), L\left(\left(\left\{p\right\}, o\right)\right) = \sigma \\
	\forall c \in conflicts(p), L\left(\left(\left\{p\right\}, c\right)\right) = \gamma
\end{align*}

We define the dependency resolution problem as mapping a set of query packages $Q$ and the resolution hypergraph $H$, to resolved graph $G$, selecting packages to satisfy dependency constraints.
Formally, given a set of packages $Q \subseteq P$ and a hyperedge-labelled directed hypergraph $H$, \textit{resolution} builds a directed graph $G$ with vertices $V(G)$ and edges $E(G)$ where,
\begin{enumerate}
	\item $G$ contains the query packages,
		\begin{wrapped}{equation*}
			\forall q \in Q, q \in V(G)
		\end{wrapped}
	\item $G$ satisfies dependencies,
        % THOMAS: as said above, I think can tweak this a bit to support CNF clauses:
        %         - remove the ! Constraint
        %         - if we really want to add some constraint, we could make sure there is only at most edge going out p with the same name -> but this won't work for Rust where you could actually depend on p.1 and p.2 in the same crate.
		% RYAN: as discussed we do need the uniqueness constraint
		\begin{wrapped}{equation*}
			\forall p \in V(G), \forall d \in deps(p), \exists! e \in d : (p, e) \in E(G) \land e \in V(G)
		\end{wrapped}
	If a package $p$ is in $V(G)$ and $d$ is a dependency of $p$; then exactly one package $e$, that satisfies $d$, is in an edge from $p$ to $e$; and $e$ is also in $V(G)$.
	\item $G$ satisfies optional dependencies,
	% $$\forall p \in V(G), \forall o \in opts(p), \forall e \in o, e \in V(G) \implies (p, e) \in E(G)$$
	% Meaning if a package $e$ that satisfies an optional dependency $o$ of package $p$ is in $V(G)$ then $p$ must have an edge to $e$.
	% TODO is this such that : notation right
    % THOMAS: why do we need `!` ? Isn't this covered by (4) ?
	% RYAN: in the case that e.g. an optional dependency is on multiple versions both of which are in the graph, where they don't conflict
	$$\forall p \in V(G), \forall o \in opts(p), (\exists e \in o : e \in V(G)) \implies (\exists! e \in o, (p, e) \in E(G))$$
	If package $p$ has its optional dependency $o$ satisfied by a package in $V(G)$, then there is exactly one package $e$ that satisfies that dependency in an edge from $p$ to $e$.
	\item $G$ satisfies conflicts,
	$$\forall p \in V(G), \forall c \in conflicts(p), p \in V(G) \implies (c \cap V(G) = \emptyset)$$
	If package $p$ is in $V(G)$ then no package $p$ conflicts with is in $V(G)$.
\end{enumerate}

An illustrated hypergraph is shown in figure~\ref{fig:hypergraph-to-graph}, which is resolved to graph $G$ according to these dependency constraints.
Formally, hypergraph $H$ has $\forall r \in R, L(r)=\delta$ and,
% $P=\{A1, B1, C1, D1, D2, D3\}$,
\begin{equation*}
\begin{gathered}
R=\{
	(\{A1\},\{B1\}),
	(\{A1\},\{C1\}),
	(\{B1\},\{D1,D2\}),
	(\{C1\},\{D2,D3\})
\}\\
\end{gathered}
\end{equation*}
This hypergraph is resolved, with $Q=\{A1\}$ to a graph $G$ with
% $V(G)=\{A1, B1, C1, D2\}$ and
\begin{equation*}
E(G)=\{
	(A1,B1),
	(A1,C1),
	(B1,D2),
	(C1,D2)
\}
\end{equation*}

% THOMAS: might be interesting to talk about the diamond problem here and say that our model do not preclude it (as some languages like Rust support it)
% RYAN: ah, I do get on to this in \ref{sec:single-versions}.

\subsection{NP-completeness}
\label{sec:complexity}

We define the \textsc{DependencyResolution} decision problem as whether a $G$ exists for a given $Q$ and~$H$.
We can show the \textsc{DependencyResolution} decision problem is NP-complete by proving that it is in NP and that it is NP-hard.
First, \textsc{DependencyResolution} is in NP as it has a proof, $G$, which can be verified in polynomial time by checking that $G$,
\begin{enumerate*}
	\item contains the query packages,
	\item satisfies dependencies,
	\item satisfies optional dependencies,
	\item satisfies conflicts.
\end{enumerate*}
This requires iterating over every vertex and edge in $G$, so can be done in polynomial time.
% RYAN: do we need a more elaborate proof?

Second, we can prove that \textsc{DependencyResolution} is NP-hard by a polynomial time reduction from the \textit{boolean satisfiability problem} (\textsc{SAT}), which is well-known to be NP-complete~\cite{cookSAT}.
The \textsc{SAT} instance is defined as a formula $F$ in conjunctive normal form (CNF) with $x$ variables and $y$ clauses.
To construct an instance of \textsc{DependencyResolution} from an instance of \textsc{SAT},
\begin{enumerate}
	\item
	For each variable we define two packages representing true and false value assignments,
	$\left(n_i, TRUE\right), \left(n_i, FALSE\right) \in P \mid 1 \leq i \leq x$ where $n_i \in N$ and $\left\{FALSE, TRUE\right\} = V_{n_i}$.

	The two packages representing true and value assignments for the same variable conflict,
	$$conflicts\left(\left(n_i, TRUE\right)\right) = \left\{\left\{\left(n_i, FALSE\right)\right\}\right\}, conflicts\left(\left(n_i, FALSE\right)\right) = \left\{\left\{\left(n_i, TRUE\right)\right\}\right\} \mid 1 \leq i \leq x$$
	\item
	For each clause we create a package with a dependency that is the set of packages corresponding to literals in the clause, with versions satisfying the literal polarity.
	We create a package for each clause $(c_j, \epsilon) \in P \mid 1 \leq j \leq y$ where $c_j \in N$ and $\{\epsilon\} = V_{c_j}$.
	Let the $k^{\text{th}}$ literal in this clause have variable $n_{j,k}$ and let $l_j$ denote the number of literals in clause $c_j$.
	Let $b_{j,k}$ represent the polarity of the literal $k^{\text{th}}$ literal, where $b_{j,k}=TRUE$ if the literal is positive and $b_{j,k}=FALSE$ if the literal is negative.
	We define,
	$$deps\left(c_j\right) = \left\{ \left\{ \left(v_{j,k}, b_{j,k}\right) \mid 0 \leq k \leq l_j \right\} \right\} \mid 1 \leq j \leq y$$
	\item We define $Q = \{q\}$ as the query package set, with $deps(q) = \{ \{c_j\} \mid 1\leq j\leq y\}$, which represents the formula $F$.
\end{enumerate}
Note that we express disjunctions in clauses with multiple packages that can satisfy a single dependency, and conjunctions between clauses with multiple dependencies.

% THOMAS: also the important part in that proof is to show that the encoding is done in linear space
% RYAN: technically not exceeding polynomial space IIRC, completed below
The reduction process can be performed in polynomial time.
Creating packages and package conflicts for $x$ variables takes $O(x)$ time.
Creating packages for $y$ clauses takes $O(y)$ time.
And creating dependencies for each clause takes time proportional to the numbers of literals in the clause.
Letting $l$ denote the total number of literals across all clauses, then creating dependencies for all clauses takes $O(l)$ time.
The total time complexity for the reduction is $O(x + y + l)$, which is classified as polynomial time.

Finding a package set that satisfies all dependencies is equivalent to finding a truth assignment that satisfies all clauses, therefore the reduction holds and \textsc{DependencyResolution} is NP-hard.
Since \textsc{DependencyResolution} is both in NP and NP-hard, it is NP-complete.

\subsection{Resolution with SAT Solvers}
\label{sec:sat-solving}

% THOMAS: you are not the first one to translate such a dependency problem into SAT. What do opam/CUDF do? Or 0install? Do they follow the same encoding? Need citations
% RYAN: TODO
We express a \textsc{SAT} instance as an \textsc{DependencyResolution} instance to prove that it is NP-hard, but we can also express a \textsc{DependencyResolution} instance as a \textsc{SAT} instance to perform dependency resolution with a SAT solver.
First, for each package we create a Boolean variable $\forall p \in P, \exists X_p$; and for each package that satisfies a dependency we create a boolean variable $\forall (p, d) \in D, \forall e \in d, \exists X_{(p,e)}$.
We then create a \textsc{SAT} instance in conjunctive normal form containing:
\begin{enumerate}
	\item Clauses to select the query packages $\forall q \in Q, (X_q)$.
	\item A clause that states if a package is chosen to satisfy a dependency then that package must be selected,
			$$\forall p \in P, \forall d \in deps(p), \forall e \in d, \left(\neg X_{(p,e)} \lor X_{e}\right)$$
		For each dependency of each package, a clause that states if the package is selected at least one package must satisfy that dependency,
			$$\forall p \in P, \forall d \in deps\left(p\right), d = \left\{ e_1, e_2, ... e_{\lvert d\rvert} \right\}, \left(\neg X_p \lor X_{(p,e_1)} \lor X_{(p,e_2)} \lor X_{(p,e_3)} \lor ... \lor X_{(p,e_{\lvert d\rvert})}\right)$$
		And a set of clauses to ensure that at most one element is chosen to satisfy the dependency,
			$$\forall p \in P, \forall d \in deps(p), \forall e_1, e_2 \in d \text{ where } e_1 \neq e_2, (\neg X_{(p, e_1)} \lor \neg X_{(p, e_2)})$$
	\item
		For each optional dependency of each package, a set of clauses that state if a package that satisfies the optional dependency is selected, at least one package must satisfy the optional dependency,
			% for all e,
			% (X_p \land X_e) implies (X_{(p,e_1}) \lor X_{(p,e_2}) \lor ... \lor X_{(p,|d|}))
			% = \neg (X_p \land X_e) \lor (X_{(p,e_1}) \lor X_{(p,e_2}) \lor ... \lor X_{(p,|d|}))
			% = (\neg X_p \lor \neg X_e) \lor (X_{(p,e_1}) \lor X_{(p,e_2}) \lor ... \lor X_{(p,|d|}))
			% = (\neg X_p \lor \neg X_e \lor X_{(p,e_1}) \lor X_{(p,e_2}) \lor ... \lor X_{(p,|d|}))
			% TODO is this o = notation right?
			$$\forall p \in P, \forall o \in opts, \forall e \in o, o = \left\{ e_1, e_2, ... e_{\lvert d\rvert} \right\}, \left(\neg X_p \lor \neg X_e \lor X_{(p,e_1)} \lor X_{(p,e_2)} \lor X_{(p,e_3)} \lor ... \lor X_{(p,e_{\lvert d\rvert})}\right)$$
		A set of clauses that state if a package satisfies an optional dependency then it must be selected,
		% i.e. for X_{(p,e}} to be true X_e must be true
		% this is equivalent to X_{(p,e)} implies X_e
		% but this does not effect the satisfiability criteria.
		% i.e. we can include an optional dependency that is not explicitly required and it would not break the satisfiability.
		% It should be a sensible heuristic of any algorithm to not include an optional dependency unless it is depended upon though
		% maybe we could include a clause saying something like an optional dependency can only be included if it is depended upon by something else?
			$$\forall p \in P, \forall o \in opts, \forall e \in o, \left(\neg X_{(p,e)} \lor X_e \right)$$
		And a set of clauses to ensure that at most one element is chosen to satisfy the optional dependency,
			$$\forall p \in P, \forall o \in opts(p), \forall e_1, e_2 \in d \text{ where } e_1 \neq e_2, (\neg X_{(p, e_1)} \lor \neg X_{(p, e_2)})$$
	\item A set of clauses that ensure conflicting packages are not present,
			$$\forall p \in P, \forall c \in conflicts(p), \forall e \in c, \left(\neg X_{p} \lor \neg X_{e}\right)$$
\end{enumerate}

% THOMAS: you just talk about soundness. What about completeness? Also is this encoding efficient? No need for a formal proof (even if that would be even better :p), but at least we need some intuition as of why it's working.
% RYAN: DONE

We can extract $G$ from the \textsc{SAT} solution, where $X_p \Leftrightarrow p \in V(G)$, and $X_{p,d} \Leftrightarrow (p, e) \in E(G)$.
By our construction of the \textsc{SAT} instance this means $G$ satisfies the \textsc{DependencyResolution} constraints~(\S\ref{sec:hyperres}).
Hence, if there is a satisfying assignment for the \textsc{SAT} instance, there is a valid resolution in the \textsc{DependencyResolution} instance, and this reduction is sound.

Given a solution for the \textsc{DependencyResolution} instance we can extract an assignment of variables, where $p \in V(G) \Leftrightarrow X_p$, and $(p, e) \in E(G) \Leftrightarrow X_{p,d}$.
This assignment satisfies all the clauses in the CNF of the \textsc{SAT} instance by the construction method.
Hence, if there is a solution to the \textsc{DependencyResolution} instance, then there is a satisfying assignment in the \textsc{SAT} instance, and this reduction is complete.

We emphasise that this is not to designed to be an efficient SAT encoding, rather to demonstrate that this formalism can be mechanically solved.
% Other algorithms include traversing the graph, selecting packages to satisfy dependencies, and backtracking on version clashes.

\section{Modelling Package Managers}
\label{sec:modelling}

% THOMAS: I think this section should say 1/ it's possible to encode many existing package managers and 2/ this encoding is actually efficient (so avoid exponential translation if possible...). It might help to have some data on how efficient SAT solvers are in another section so we have an idea of how big/bad our encoding could be.
% RYAN: I've emphasised we're demonstrating the expressivity of our minimal model, which could be extended for more efficient solves.

In this section we will illustrate how a variety of properties of real-world package managers~(\S\ref{sec:ontology}) can be modelled in the HyperRes formalism~(\S\ref{sec:formalism}).
This formalism has been defined to be as minimal as possible to capture the common core of package management, but is expressive enough to model complex functionality such as boolean algebra~(\S\ref{sec:boolean}) and features parameterisation~(\S\ref{sec:features}).
% These encodings may not lead to an efficient \textsc{SAT} instance~(\S\ref{sec:sat-solving}), but in practice, the HyperRes formalism could be extended to support this functionality as a first-order property.
% The key point is that there is a common core to dependency resolution which can be expressed using our minimal formalism.

\subsection{APT}

APT is the package manager for the Debian Linux distribution (Table~\ref{tbl:comparison}).
It manages system-wide packages, ensuring that dependencies are met and conflicts are avoided.
It handles a single version of a package at a time, and supports virtual packages which allow multiple packages to provide the same functionality.

% Build/dev/runtime
% Binary packages creating multiple source packages - dynamic dependencies

\begin{wrapfigure}[18]{R}{0.25\textwidth}
	\captionsetup{justification=centering}
	\centering
	\includegraphics[width=\linewidth]{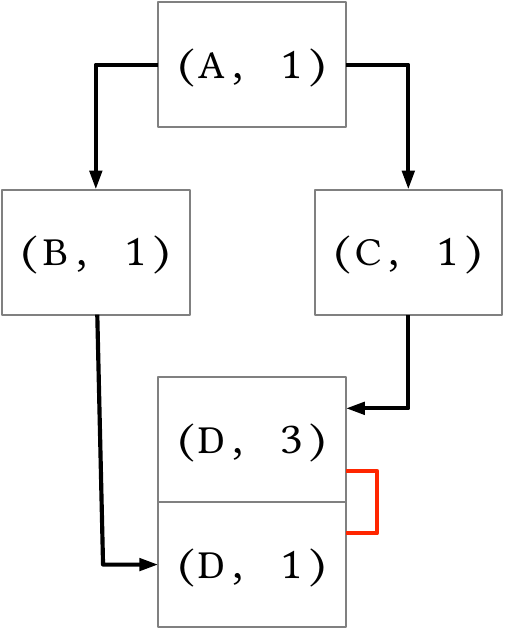}
	\caption{A resolution hypergraph $H$ exhibiting the `diamond problem'.}
	\Description{A resolution hypergraph $H$ where,
			$P=\{A1, B1, C1, D1, D3\}$,
			$D=\{
				(\{A1\},\{B1\}),
				(\{A1\},\{C1\}),
				(\{B1\},\{D1\}),
				(\{C1\},\{D3\})
			\}$,
			$C=\{
				(\{D1\},\{D3\}),
				(\{D3\},\{D1\})
			\}$,
			$R=D \cup C$,
			$\forall r \in D, L(r)=\delta$, and
			$\forall r \in C, L(r)=\gamma$.
	}
	\label{fig:diamond-hypergraph}
\end{wrapfigure}

\subsubsection{Single Versions}
\label{sec:single-versions}
As with many packages managers APT only supports deploying a single version of a package name at a time~(\S\ref{sec:table:concurrent-versions}).
We can encode this in the HyperRes formalism by saying for each package $p$, $conflicts(p)$ contains every other package with the same package name,
\begin{wrapped}{equation*}
	\forall (n, v) \in P, \left\{\left(n, v'\right) \mid v' \in V_n, v' \neq v\right\} \subseteq conflicts\left(\left(n, v\right)\right)
\end{wrapped}
This means that for all package names in $V(G)$, there exists exactly one version for each,
\begin{wrapped}{equation*}
	\forall (n, v) \in V(G), \forall (n', v') \in V(G), n = n' \implies v = v'
\end{wrapped}

Consider a hypergraph as depicted in Figure~\ref{fig:diamond-hypergraph}, where only a single version of a package name is allowed, as encoded as a conflict between \texttt{D1} and \texttt{D2}.
Note that for diagramming simplicity we represent this as an undirected `conflict set' where every package connected to the solid line conflicts with all the others.
This hypergraph exhibits the `diamond dependency problem', so called as the graph with edges $\{(A,B), (A,C), (B,D), (C,D)\}$ forms a diamond, which has no solution.
Formally, $H$ has,
%$P=\{A1, B1, C1, D1, D3\}$,
\begin{equation*}
\begin{gathered}
D=\{
	(\{A1\},\{B1\}),
	(\{A1\},\{C1\}),
	(\{B1\},\{D1\}),
	(\{C1\},\{D3\})
\}\\
C=\{
	(\{D1\},\{D3\}),
	(\{D3\},\{D1\})
\}\\
\forall r \in D, L(r)=\delta \quad
\forall r \in C, L(r)=\gamma \quad
R= D \cup C
\end{gathered}
\end{equation*}

\subsubsection{Version Ordering}
\label{sec:version-ordering}
Like many package managers APT imposes a total lexicographical ordering of versions~\cite{debianVersions}.
For example, version \texttt{2.0} of a package will be greater than version \texttt{1.0}.
The dependency resolution problem does not specify a preference, but an algorithm to solve the problem can take this into account by trying packages with higher versions first.
For example, if we create literals in a \textsc{SAT} clause~(\S\ref{sec:sat-solving}) according to this ordering and try literals in the order they appear in the clause, we will prefer higher versions.
% TODO cite https://roscidus.com/blog/blog/2014/09/17/simplifying-the-solver-with-functors/#optimising-the-result
% TODO consider `recommended' packages and cost functions like OPIUM's

\subsubsection{Upgrades}

\begin{wrapfigure}{R}{0.25\textwidth}
	\captionsetup{justification=centering}
	\centering
	\includegraphics[width=\linewidth]{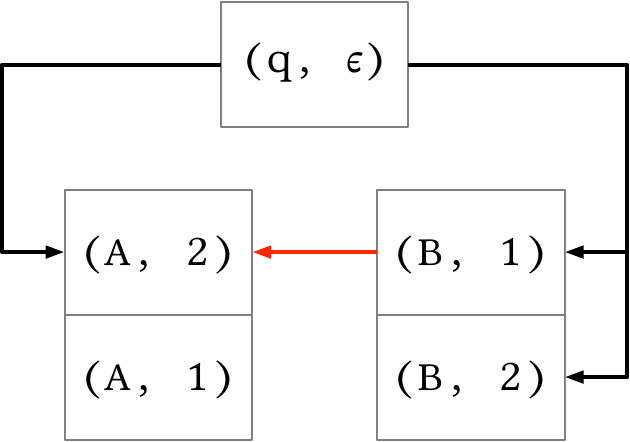}
	\caption{A resolution hypergraph expressing an upgrade.}
	\Description{A resolution hypergraph $H$ where,
			$P=\{q, A1, A2, B1, B2\}$,
			$D=\{
				(\{q\},\{A1, A2\}),
				(\{q\},\{B1, B2\}),
			\}$,
			$C=\{
				(\{B1\},\{A2\}),
			\}$,
			$R=D \cup C$,
			$\forall r \in D, L(r)=\delta$, and
			$\forall r \in C, L(r)=\gamma$.
	}
	\label{fig:upgrade}
\end{wrapfigure}

The feedback mapping in Figure~\ref{fig:pipeline} shows an input from the resolved graph into the resolution graph, but the HyperRes formalism doesn't take the existing state of the system into account.
HyperRes doesn't consider how package managers like APT are used to upgrade individual packages.
We can model this by pinning installed versions in our query set.

Consider a scenario where we have updated the Debian repository~(\S\ref{sec:table:repository}) and want to upgrade a single package to a new version.
We can represent this in our formalism by creating a virtual package that has a dependency set containing the versions greater than the current versions of the package name being upgraded, as well as a dependency on every other installed package.
To deploy the upgrade we can take the difference from the resulting resolved graph and the pre-upgrade resolved graph, removing and adding packages as appropriate.
This doesn't consider a scenario where the upgraded package and existing packages can't co-exist.

If there's a conflict with the versions of the installed packages, we can depend on their greater versions.
That is, for each installed package name, add a dependency set on the current version and any greater versions to the virtual package.
Figure~\ref{fig:upgrade} shows a resolution hypergraph with a virtual package $q$ where $Q=\{q\}$ expressing an upgrade on package $A1$ with an installed package $B1$.
The installed package $B1$ conflicts with the new version of $A$, so $B1$ will also need upgrading.
We could express a preference for minimising upgrades with a ordering of these installed package versions preferring lower versions~(\S\ref{sec:version-ordering}).

The Common Upgradeability Description Format (CUDF)~\cite{cudf} provides a standard format for expressing such upgrade problems that can be solved by various algorithms.

% The feedback mechanism is described as optional as some package managers and solvers don't take the current state of the system into account, leading to reproducible solves at the cost of minimal upgrades.

\subsubsection{Virtual Packages}
APT packages can express dependencies on virtual packages that other packages `provide'~\cite{debian}.
For example, both \texttt{openssh-server} and \texttt{dropbear-bin} provide the virtual package \texttt{openssh-server}.
We can encode a dependency on a virtual package in the resolution hypergraph with a dependency on the set of all packages which provide it, as depicted in Figure~\ref{fig:packulus-miraculous}~(\S\ref{sec:illustration:different-names}).
Assuming a set of virtual packages $P_v$ and a function which contains the set of virtual packages a provided by a package, $provides : P \to 2^P$, we can formally define this encoding with,
$$\forall p_v \in P_v, deps(p_v)=\{\{p \mid p_v \in provides(p), \forall p \in P \}\}$$

% THOMAS: I think that some of those virtual packages has a notion of 'default' package. Unclear how to encode this choice here
% RYAN: perhaps as a version ordering \ref{sec:version-ordering}?
% RYAN: I think they also use a disjunction sometimes, i.e. (openssh-server | ssh-server)

% https://doc.opensuse.org/projects/satsolver/11.2/
% A provides B -> B == A (replace all occurrences of B in CNFs with A)

\subsubsection{Computer Architecture}
\label{sec:architecture}

\begin{wrapfigure}{r}{0.333\textwidth}
	\captionsetup{justification=centering}
	\centering
	\vspace{-1em}
	\includegraphics[width=\linewidth]{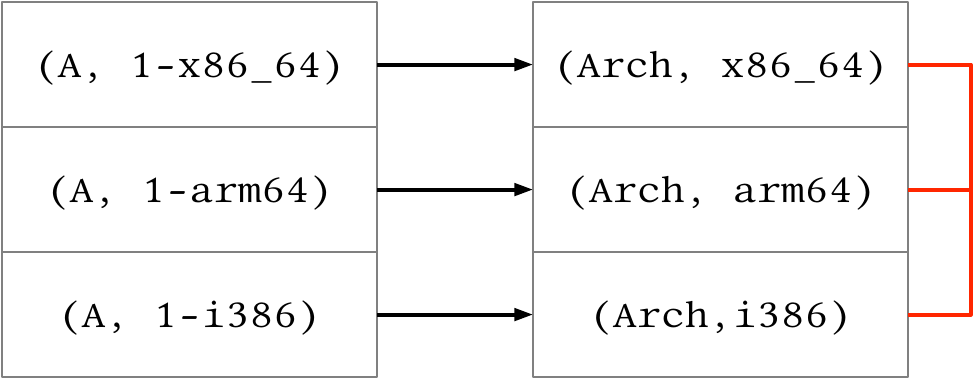}
	\caption{Encoding computer architecture in the resolution hypergraph.}
	\Description{A resolution hypergraph $H$ where,
			$P=\{A.1-x86_64, A.1-ARM64, A.1-i386, Arch.x86_64, Arch.ARM64, Arch.i386\}$,
			$D=\{
				(\{A.1-x86_64\},\{Arch.x86_64\}),
				(\{A.1-ARM64\},\{Arch.ARM64\}),
				(\{A.1-i386\},\{Arch.i386\}),
			\}$,
			$C=\{
				(\{Arch.x86_64\},\{Arch.ARM64\}),
				(\{Arch.x86_64\},\{Arch.i386\}),
				(\{Arch.ARM64\},\{Arch.x86_64\}),
				(\{Arch.ARM64\},\{Arch.i386\}),
				(\{Arch.i386\},\{Arch.ARM64\}),
				(\{Arch.i386\},\{Arch.x86_64\}),
			\}$,
			$R=D \cup C$,
			$\forall r \in D, L(r)=\delta$, and
			$\forall r \in C, L(r)=\gamma$.
	}
	\label{fig:arch}
\end{wrapfigure}

Packages managers that provide binary packages often provide different versions of a package for different architectures, and APT is no exception.
APT points to different package versions depending on the architecture of the machine it is running on.
We can encode computer architecture as a package in the resolution hypergraph in order to support resolving for a particular architecture.

We create a name $\texttt{arch}$ in $N$.
The versions of this name will be the enumeration of supported architectures, e.g.
$$\{\texttt{x86\_64}, \texttt{ARM64}, \texttt{i386} \} = V_\texttt{arch}$$
We define conflicts between all the architecture versions to force a selection,
$$\forall v \in V_\texttt{arch}, \left\{\left(\texttt{arch}, v'\right) \mid v' \in V_\texttt{arch}, v' \neq v\right\} \subseteq conflicts((\texttt{arch}, v))$$
The associated architecture for a debian package can be added as a dependency.
Figure~\ref{fig:arch} shows a package \texttt{A} with versions \texttt{1-x86\_64}, \texttt{1-ARM64}, and \texttt{1-i386} depending on the appropriate \texttt{arch} version.
By including an \texttt{arch} package in the query set a resolve can be restricted according to the hardware available.
% TODO example of a package that might require specific hardware?

\subsection{Opam}

Opam is the package manager for the OCaml programming language (Table~\ref{tbl:comparison}).
It supports complex dependency formulae with global and local variables, and requires an acyclic resolved dependency graph to perform build steps in an order such that all the dependencies of a package are built before it is built.

\begin{wrapfigure}[14]{R}{0.25\textwidth}
	\captionsetup{justification=centering}
	\centering
	\includegraphics[width=\linewidth]{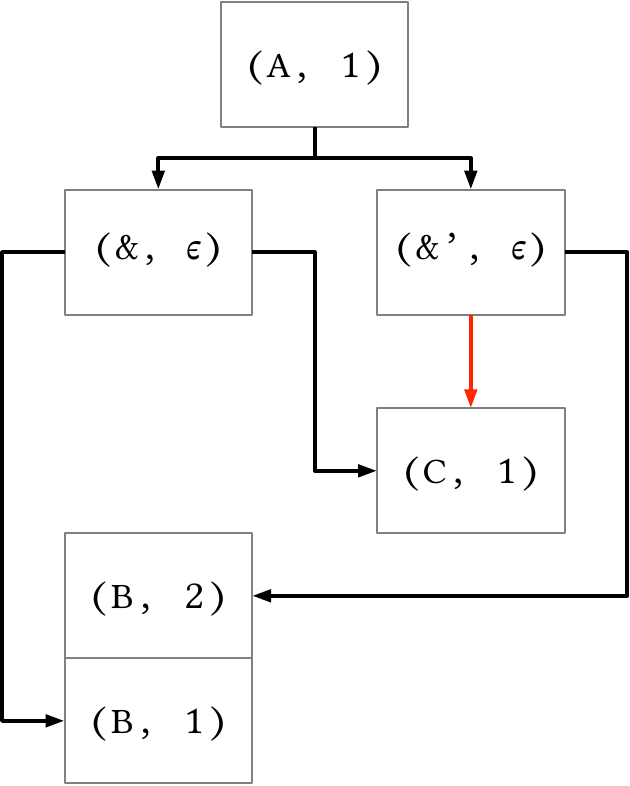}
	\caption{Encoding boolean algebra in a hypergraph.}
	\Description{A resolution hypergraph $H$ where,
			$P=\{A1, \&\epsilon, \&'\epsilon, B1, C1, B2\}$,
			$D=\{
				(\{A.1\},\{\&\epsilon, \&'\epsilon\}),
				(\{\&\epsilon\},\{B1\}),
				(\{\&\epsilon\},\{C1\}),
				(\{\&'\epsilon\},\{B2\}),
			\}$,
			$C=\{
				(\{\&'\epsilon\},\{C1\}),
			\}$,
			$R=D \cup C$,
			$\forall r \in D, L(r)=\delta$, and
			$\forall r \in C, L(r)=\gamma$.
	}
	\label{fig:boolean}
\end{wrapfigure}

% dev dependencies

% Flags
% https://opam.ocaml.org/doc/Manual.html#opamfield-flags

\subsubsection{Dependency Formula}

% THOMAS: with the new encoding scheme, I would just CNF encode the dependency formula (can be done in polynomial space) and that's it.
% RYAN: we haven't gone for this new encoding scheme due to the requirement of choosing only one package to satisfy a dependency
% THOMAS: the negation: encoding the negation of a formula as conflict will not work; you want CNF transform + manually negate the version set of packages; i.e. n { != 3 } is n { = 1 || = 2 || n = 4} (which again is polynomial)
% RYAN: yes... I was confusing the package formula and version filters
% RYAN: actually I've gone back on this -- see below

\paragraph{Boolean Algebra}
\label{sec:boolean}

Opam supports `package formula' and `version formula', boolean algebra of packages names and package versions respectively~\cite{opam}.
As we have shown~(\S\ref{sec:complexity}) we can encode a \textsc{SAT} expression in our HyperRes formalism, and can encode opam's package formulae in the same way.
Disjunctions can be represented by a dependency set, conjunctions with a `virtual' package that depends on both the operands of the conjunctions, and negations with conflicts.
% Negation can only be expressed in version formula (conflicts are used to negate package names), for example \verb|"package-a" { ! "1.0.0" }|, which can be represented by the set of packages not matching the negation, e.g. \texttt{0.0.1} and \texttt{2.0.0}.
% RYAN: negations as conflicts is simpler, even if expanding versions might be more efficient

Figure~\ref{fig:boolean} demonstrates how boolean algebra can be represented in HyperRes with a dependency from A1 that corresponds to the boolean expression, $(B1 \land C1) \lor (\neg C1 \land B2)$.
Note that the $\&$ symbol represents a virtual package with an empty version $\epsilon$ and a globally unique name.
% RYAN: TODO Formalise this

\paragraph{Variables}
Opam also supports global and package-local variables in its package formulae.
With the global \texttt{os-distribution} variable in a package formula, the formula's dependency will be conditional on the operating system the package manager is deploying the package on (\S\ref{sec:illustration:system-deps}).
%When used to parameterise build and install scripts, we can consider it part of package language evaluation~(\S\ref{sec:pipeline:parsing}).
We can encode this in resolution hypergraph as we did with computer architecture for APT~(\S\ref{sec:architecture}).
That is, we create a package \texttt{os-distribution} with versions enumerating the possible operating system distributions.
For a package dependent upon \verb|["gcc"] {"os-distribution" = "debian"}| we can depend on a virtual package which itself depends on the \texttt{os-distribution} package with version \texttt{debian}, and a version of \texttt{gcc}.
% RYAN: TODO Formalise this

Package-local variables can be encoded as optional dependencies.
The package-local \texttt{with-test} denotes variables that are only required if tests have been enabled for the package.
For example, consider a package \texttt{(polars, 0.1.0)} that has a dependency on \texttt{"alcotest" \{with-test\}}.
We can create a virtual package \texttt{(polars, (0.1.0, test))} which \texttt{(polars, 0.1.0)} optionally depends on, where \texttt{(polars, (0.1.0. test))} depends on \texttt{alcotest}.
To enable tests for \texttt{(polars, 0.1.0)} the package \texttt{(polars, (0.1.0. test))} can be installed.
% An example of another variable which can be encoded in this mechanism is the \texttt{build} variable.

\subsubsection{Acyclic Resolved Graph}

Opam has an additional requirement for the resolved graph over those specified in section~\ref{sec:hyperres}, that it must be \textit{acyclic}.
As part of its deployment~(\S\ref{sec:pipeline:deployment}) opam performs a topological sort of the graph --- which is only possible with a directed acyclic graph --- to assemble a list of actions to be executed sequentially.
These actions build packages, and if all the required dependencies of a package aren't provided prior to its build, the build will fail.
An exception to this is packages marked with the \texttt{post} variable, which denotes where the cycle can be broken as a dependency isn't a build-time dependency.
Other package managers deal with cycles by breaking them at some arbitrary point, like APT with binary packages~\cite{debian}; or not performing a topological sort at all, npm lays out files and relies on language support to break cycles between modules.
% npm https://nodejs.org/api/modules.html#modules_cycles

We can formally define this as for the resolved graph $G$, for any sequence of edges $e_1, e_2, ..., e_n$ in $G$ where $e_i = (p_i, e_i)$, for $1 \leq i \leq n,$ it holds that $p_i \neq p_{i+1}$ for any $1 \leq i < n$.

\subsection{Cargo}
\label{sec:cargo}

Cargo is the package manager is the Rust language's package manager and build system (Table~\ref{tbl:comparison}).
It supports multiple versions of the same package coexisting in a project, provided they have different major versions.
It also uses features to enable or disable optional functionality in packages.

\subsubsection{Concurrent Versions}

\begin{wrapfigure}[13]{R}{0.25\textwidth}
	\captionsetup{justification=centering}
	\centering
	\vspace{-1em}
	\includegraphics[width=\linewidth]{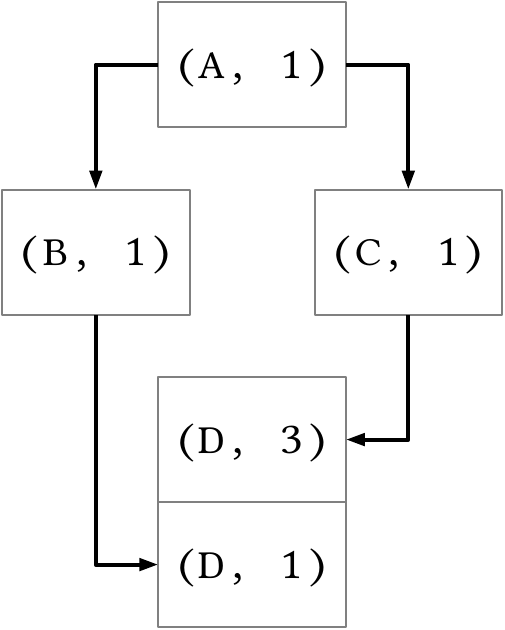}
	\caption{A multi-version resolved graph $G$.}
	\Description{A resolved graph $G$ where $q=A1$,
		$V(G)=\{A1, B1, C1, D1, D3\}$, and
		$E(G)=\{
			(A1,B1),
			(A1,C1),
			(B1,D1),
			(C1,D3)
		\}$.
	}
	\label{fig:multiversion-graph}
\end{wrapfigure}

As discussed in section~\ref{sec:table:concurrent-versions}, Cargo supports multiple versions of packages in a resolved graph.
To be precise, it only allows different major version numbers in the semantic versioning scheme~\cite{cargoResolver, semver}.
To model this in the HyperRes formalism, we can add conflicts as described in section~\ref{sec:single-versions}, but only if the packages share the same major version.
Figure~\ref{fig:packulus-miraculous} demonstrates this (\S\ref{sec:illustration:concurrent-versions}).

% Formally, assuming a function $major$ which extracts the major version from a version,
% \begin{wrapped}{equation*}
% 	\begin{gathered}
% 		\forall (n, v) \in P,\\
% 		\left\{\left(n, v'\right) \mid v' \in V_n, major(v') \neq major(v)\right\} \subseteq conflicts\left(\left(n, v\right)\right)
% 	\end{gathered}
% \end{wrapped}

This easing of the constraints allows a solution to the diamond dependency problem from Figure~\ref{fig:diamond-hypergraph}.
If we remove the conflicts set $C$, and let $R=D$, then $G$ in Figure~\ref{fig:multiversion-graph} represents a solution.

% Another way to avoid NP-completeness is to combine the previous two. As the examples already hint at, if packages follow semantic versioning, a package manager might automatically use the newest version of a dependency within a major version but then treat different major versions as different packages.
% https://research.swtch.com/version-sat

% rust even when given 1(a) it will give 2(b) instead of 1(b).
% https://github.com/rust-lang/cargo/issues/10599
% https://github.com/rust-lang/cargo/blob/master/src/cargo/core/resolver/mod.rs
% https://github.com/pubgrub-rs/pubgrub
% https://github.com/dart-lang/pub/blob/master/doc/solver.md
% Answer Set Programming https://potassco.org/book/
% https://doc.rust-lang.org/cargo/reference/resolver.html#feature-resolver-version-2

\subsubsection{Features}
\label{sec:features}
% https://doc.rust-lang.org/cargo/reference/features.html
Rust supports parametrising packages with features.
For example, the \texttt{regex} package has features to enable performance optimisations and Unicode support.
When enabled features can require additional dependencies.
Packages express features that they require when depending on a package, and when resolving dependencies packages will take the union of all the features that they are depended upon with.
We will describe an extension to the HyperRes formalism to support parametrising dependencies with features, and show how we can encode this extended formalism using versions in the original formal system.

\paragraph{Extending HyperRes to support features.}

\begin{wrapfigure}[15]{R}{0.25\textwidth}
	\captionsetup{justification=centering}
	\centering
	\includegraphics[width=\linewidth]{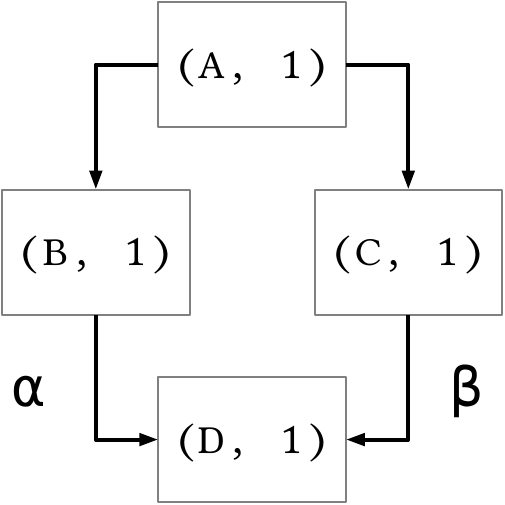}
	\caption{A resolution hypergraph where hyperedges are parameterised by features.}
	\Description{A feature-extended resolution hypergraph $H$ where,
		$V(G)=\{A1, B1, C1, D1\}$,
		$E(G)=\{
			(A1,B1),
			(A1,C1),
			(B1,D1),
			(C1,D1)
		\}$,
		$F_{D1}=\{\alpha, \beta\}$,
		$L_F\left(\left(\left\{B1\right\}, \left\{D1\right\}\right)\right)=\alpha$, and
		$L_F\left(\left(\left\{B1\right\}, \left\{D1\right\}\right)\right)=\beta$.
	}
	\label{fig:extended-features}
\end{wrapfigure}
% Ryan: TODO add depexts

% THOMAS: I haven't reviewed this in detail
To start, we define a set of package features for every package, $\forall p \in P, \exists F_p \subseteq F$ and $F$ as the set of all possible features.
Then, we define a multi-label function mapping the dependency hyperedges of the resolution hypergraph $H$ to sets of features required for the dependency.
We define this mapping $L_F~:~2^P~\times~2^P~\to~F$ where $\forall~p~\in~P,~\forall~d~\in~deps(p),~L_F((\{p\},d))$ is the set of features required by hyperedge $(\{p\},d)$.
Note that we assume the same feature set can be applied to all packages that can satisfy a dependency, $\forall p \in P, \forall d \in deps(p), \exists F_d : e \in d \implies F_e = F_d$.

We add a criterion to the dependency resolution problem, that $G$ unifies features, with a multi-label function mapping the vertices of the resolved graph $G$ to sets of features used by that package $L_V : P \to F$.
\begin{equation*}
	\forall p \in V(G), \forall d \in deps(p), \exists! e \in d : (p, e) \in E(G), L_F\left(\left(\left\{p\right\},d\right)\right) \subseteq L_V(e)
\end{equation*}
Meaning that for each package $e$ chosen to satisfy dependency $d$ of package $p$, the features selected for $e$ -- $L_V(e)$ -- must include the features required by dependency $d$ -- $L_F((p,d))$.

The features of a package can affect its dependencies, we define a function $fdeps : P \times F \to 2^{2^P}$ function which maps a package and its feature to a set of additional dependencies required by that feature.
We add another criterion of HyperRes that $G$ satisfies feature dependencies,
\begin{wrapped}{equation*}
	\begin{gathered}
		\forall p \in V(G), \forall f \in L_V(p), \forall d \in fdeps((p, f)),\\
		\exists! e \in d : (p, e) \in E(G) \land e \in V(G)
	\end{gathered}
\end{wrapped}
Meaning if a package $p$ is in $V(G)$ with features $f$ selected, and $d$ is a dependency of $p$ with features $f$; then exactly one package $e$, that satisfies $d$, is in an edge from $p$ to $e$; and $e$ is also in $V(G)$.

Figure~\ref{fig:extended-features} shows an example hypergraph where $B1$ depends on $D1$ with feature $\alpha$ and $C1$ depends on $D1$ with feature $\beta$.
Formally,
\begin{wrapped}{equation*}
% \begin{equation*}
	\begin{gathered}
		F_{D1}=\{\alpha, \beta\}, \quad
		L_F\left(\left(\left\{B1\right\}, \left\{D1\right\}\right)\right)=\alpha, \quad
		L_F\left(\left(\left\{B1\right\}, \left\{D1\right\}\right)\right)=\beta \quad
	\end{gathered}
% \end{equation*}
\end{wrapped}

\paragraph{Encoding the extended formalism in HyperRes.}
% Ryan: TODO how does this work when multiple versions are selected?

\begin{wrapfigure}[20]{R}{0.25\textwidth}
	\captionsetup{justification=centering}
	\centering
	\includegraphics[width=\linewidth]{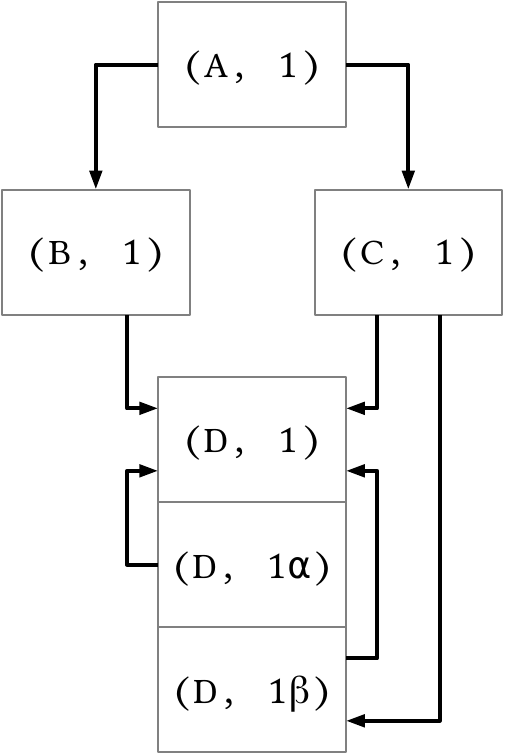}
	\caption{A resolution hypergraph where features are encoded as optional dependencies.}
	\Description{A resolution hypergraph $H$ where,
		$P=\{A1, B1, C1, D1, D1\alpha, D1\beta\}$,
		$D=\{
			(\{A1\},\{B1\}),
			(\{A1\},\{C1\}),
			(\{B1\},\{D1\}),
			(\{B1\},\{D1\alpha\}),
			(\{C1\},\{D1\})
			(\{C1\},\{D1\beta\}),
		\}$,
		$O=\{
			(\{D1\},\{D1\alpha\}),
			(\{D1\},\{D1\beta\}),
		\}$,
		$R=D \cup C$,
		$\forall r \in D, L(r)=\delta$, and
		$\forall r \in O, L(r)=\sigma$.
	}
	\label{fig:encoded-features}
\end{wrapfigure}

% THOMAS: this is a pretty intrusive encoding - might be interesting to discuss the added complexity (ie. Seems exponential in the number of features). Could we do better? Maybe using optional dependencies (like we do in opam with the `variant-*` packages)
% RYAN: the key functionality here is the feature unification. I've added a disclaimer to the start of the modelling section saying we're really showing the expressivity of the formalism but that the model could be extended for more efficient encodings.
% RYAN: I have a better idea -- using optional dependencies

To show that the HyperRes formalism is expressive enough to describe features, we will encode the extended formalism in the HyperRes formalism.
For every package name we define a `feature version' set as pairs of version and feature supported by that version,
\begin{wrapped}{equation*}
	\forall n \in N, \exists FV_n = \left\{ \left(v, s\right) \mid f \in F_{\left(n,v\right)}, v \in V_n\right\}
\end{wrapped}
Assuming an ordering of features, and a new separator, we can encode this feature version as a string.
If we had a package name \texttt{D} with version \texttt{1}, and features \texttt{a} and \texttt{b}, we could encode the new versions as \texttt{1+a} and \texttt{1+b}.

We define a new version set for each package name containing the normal versions and feature versions,
\begin{wrapped}{equation*}
	\forall n \in N, \exists V_n' = V_n \cup FV_n
\end{wrapped}
And we define a new package set,
\begin{wrapped}{equation*}
	P' = \left\{ \left(n, v\right) \mid n \in N, v \in V_n' \right\}
\end{wrapped}
We define a new dependency function $deps'$ and let,
\begin{wrapped}{equation*}
	\forall n \in N, \forall (v, f) \in FV_n, deps'\left(\left(n,\left(v,f\right)\right)\right)=\left\{\left\{\left(n,v\right)\right\}\right\}\cup fdeps((n,v), f)
\end{wrapped}
Meaning for every new feature version package we depend on the corresponding package version without the feature version; as well as any new dependencies introduced by said feature.
So package \texttt{(D,1+a)} would depend on package \texttt{(D,1)}.
We also let,
\begin{equation*}
	\forall p \in P, deps'\left(p\right)= deps(p) \cup \left\{\left\{\left(n, \left(v,f\right)\right) \mid \left(n, v\right) \in d\right\} \mid \forall f \in L_F\left(\left(\left\{p\right\},d\right)\right), d \in deps\left(p\right)\right\}
\end{equation*}
Meaning where we depended on a package $p$, we now also depend on the versions representing the features that we depended on $p$ with.
If we depended on a package with \texttt{D} with version \texttt{1} and feature \texttt{a}, we would now depend on versions \texttt{1} and \texttt{1+a}.

We construct the hypergraph as before with $P'$ in place of $P$ and $deps'$ in place of $deps$.
Figure~\ref{fig:encoded-features} shows an example of an encoding of Figure~\ref{fig:extended-features} with this scheme.
Not that we omit separators between versions and features for the sake of brevity.
% Ryan: TODO diagram this

With a resolution to a resolved graph $G'$, we can extract the feature set of packages in $P$ with,
$$\forall n \in N, \forall (v, f) \in VF_n, (n, (v, f)) \in V(G') \implies f \in L_V\left(\left(n, v\right)\right)$$
And the dependencies introduced by feature selection with,
$$\forall n \in N, \forall (v, f) \in VF_n, \left\{\left(n, \left(v, f\right)\right), d\right\} \in E(G') \implies \left(\left\{\left(n, v\right)\right\}, d\right) \in E(G)$$
% This encoding is in fact more expressive than the extended formalism, as we can express conflicts on particular sets of features.
% RYAN: TODO formally prove soundness and completeness

\subsection{Nix}
\label{sec:Nix}

Nix is a package manager that pioneered the functional model of software deployment~\cite{dolstraNixSafePolicyFree2004}.
It describes packages in Nix DSL expressions which are `translated' with \texttt{nix-instantiate} to store derivations, a JSON file describing the inputs, outputs, and build steps of a derivation (package).
Store derivations are `realised' into files in the Nix store with \texttt{nix-store --realise}.
In our package management pipeline the translation is modelled as evaluation of the packaging language to bundle format, and realisation as deployment of the bundle formats to the filesystem (Table~\ref{tbl:comparison}).

% TODO: talk about how things like opam-nix can do a solve and output nix DSL derivations, but this doesn't allow cross-ecosystem dependencies

\subsubsection{Specific dependencies}

The store derivation format of Nix does not support expressing a set of possible packages that can satisfy a dependency, instead pointing to the exact package required.
We can model this in the resolution hypergraph by restricting the codomain to a size of one,
$$\forall r \in R, r = \left(\left\{p\right\}, \left\{e\right\}\right) : p \in P \land e \in P$$

% It would be possible to perform dependency resolution by writing a solver in the Nix DSL, which would mean resolution happens \textit{in} the evaluation step in our pipeline, but Nix does not support this functionality.

Nix also does not support optional dependencies or conflicts $\forall r \in R, L(r)=\delta$.
Instead of optional dependencies, derivations are generated by functions in the Nix DSL which can be parameterised to enable varying functionality.
And as Nix stores components at paths containing unique cryptographic hashes, it allows multiple packages to co-exist that would otherwise conflict with each other~(\S\ref{sec:table:concurrent-versions}).
% Pname
% version
% attrname
This turns the dependency resolution problem to a simple walk of the tree where satisfying dependencies means,
% $$E(G) \subseteq \left\{\left(p,e\right) \mid \left(\left\{p\right\}, \left\{e\right\}\right) \in R \right\}$$
$$\forall p \in V(G) : deps(p) = \{e\}, (p, e) \in E(G) \land e \in V(G)$$

%NB the dependency resolution problem is a decision problem and not turing-complete

% \subsection{Unison}
%
% Unison Code Manager
%
% No name conflicts
% at language level instead of build system name mangling like cargo or file paths like Nix

\section{Bridging Ecosystems}
\label{sec:ecosystem}

Having established that HyperRes~(\S\ref{sec:formalism}) can model a wide array of package managers individually~(\S\ref{sec:modelling}), we now turn our attention to the bigger picture of how to unify these ecosystems by resolving dependencies across them all.
We aim to enable projects that span multiple languages and platforms to manage their dependencies coherently, overcoming the current fragmentation caused by incompatible package management systems.

We thus extend the HyperRes formalism to support cross-ecosystem dependencies (\S\ref{sec:ecosystem:formalism}), demonstrate how to resolve such dependencies (\S\ref{sec:ecosystem:resolving}), and outline how translating packages bidirectionally between ecosystems is now possible (\S\ref{sec:ecosystem:translation}).

\subsection{HyperRes for Multiple Ecosystems}
\label{sec:ecosystem:formalism}

Our early example in Figure \ref{fig:packulus-miraculous} shows a resolution hypergraph including packages from multiple different ecosystems, with some dependencies transcending these boundaries.
To introduce this notion of ecosystem to the HyperRes formalism, we can add an ecosystem vertex labelling to distinguish packages belonging to different ecosystems.

We define a set of ecosystems $T$, with a package namespace for each $\forall t \in T, \exists N_t$.
We define the set of names $N=\{n \mid n \in N_t, t \in T\}$, and a package set for each ecosystem using the version sets as previously defined $\forall t \in T, P_t=\left\{\left(n, v\right) \mid v \in V_n, n \in N\right\}$.
And we define an ecosystem labelling function, $L_e : P \to T$ as,
$\forall t \in T, \forall p \in P_t, L_e(p)=t$.
Cross-ecosystem relationships are relationships where,
$\left(\{p\},\left\{d\right\}\right) \in R, \exists e \in d : L_e(p) \neq L_e$.

We distinguish between ecosystems with this labelling as they differ in how their metadata bundles are parsed to create a hypergraph (the `on-ramp' to the formalism from the ecosystem), and how packages in the resolved graph are provided (the `off-ramp' to the formalism to invoke the lower-level installer for that ecosystem).

This allows HyperRes graphs to be resolved in one pass {\em across} ecosystems with \textit{no modification} to HyperRes aside from this additional labelling.
In this cross-ecosystem HyperRes, we can include multiple repositories in any ecosystem modelled by packages.
This lets us, for example, determine which version of Debian Linux would provide the required version of a language dependency (\S\ref{sec:illustration:repository-selection}).
We can also include the default Linux kernel version provided by various repository versions as a cross-ecosystem package (\ref{sec:illustration:system-deps}), allowing for low-level dependencies or GPU drivers to be resolved reliably, and percolate these constraints higher up the stack into language-specific ecosystems such as Python's.
We can further model constraints such as computer architecture (\S\ref{sec:architecture}) to select for optimised or hardware-dependent solutions, or do the opposite and scan for a portability matrix for a given software release.

\subsection{Cross-Ecosystem Resolution}
\label{sec:ecosystem:resolving}

\newcommand\sqr[1]{{\color{#1}{\ding{110}}}}
\definecolor{opam}{HTML}{bfffbf}
\definecolor{cargo}{HTML}{ffbfbf}
\definecolor{deb}{HTML}{bfffff}

\begin{figure}[p]
	\centerline{
		\includegraphics[height=\textheight]{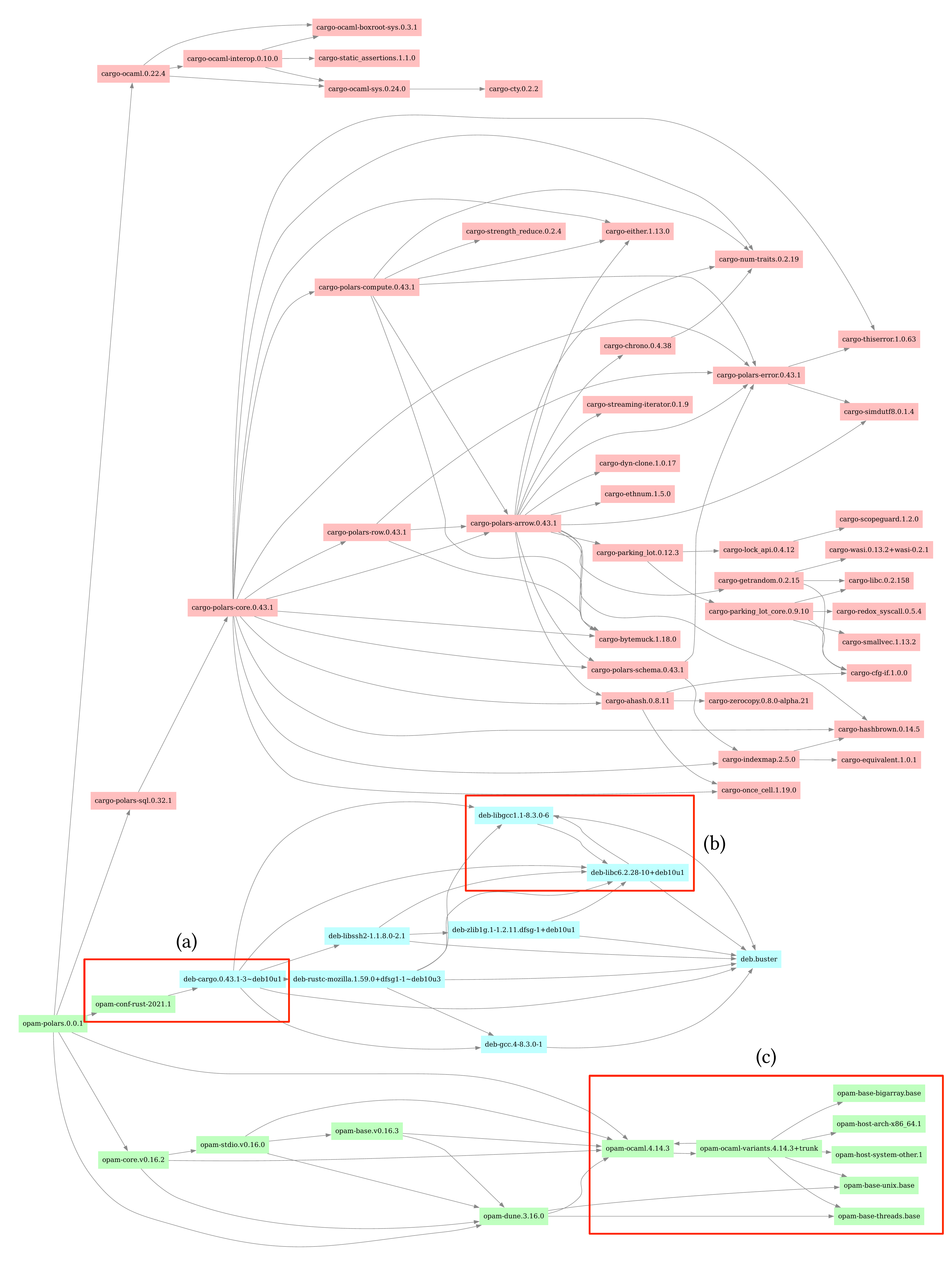}
	}
	\captionsetup{justification=centering}
	\caption{A subgraph of the full cross-ecosystem resolved graph querying opam's \texttt{polars.1.0.0} library.\\(a) shows a Rust to Debian dependency, (b) shows a Debian cycle, (c) shows an OCaml post-cycle.\\
		\sqr{cargo} Cargo \sqr{deb} Debian \sqr{opam} Opam
	}
	\Description{A subset of the resolved graph of a cross-ecosystem dependency on opam's \texttt{polars.1.0.0} library.}
	\label{fig:solve}
\end{figure}

We have implemented a number of translations from ecosystem-specific metadata bundles to a prototype HyperRes encoding.
For some package managers, like APT and APK, this involves parsing and translating static metadata.
For others with tighter integration between their packaging language and metadata bundles, like opam and Cabal (\S\ref{sec:pipeline:parsing}), OCaml and Haskell libraries were used to interpret opam and Cabal files to an in-memory representation that was translated to the HyperRes encoding.
We currently have translations from APK, Cabal, Cargo, CentOS, Debian, Fedora, Homebrew, opam, OpenSUSE, and Pacman; totalling just over 1.7 million packages.

While none of these ecosystems natively supports cross-ecosystem resolving, we can add these ecosystem-transcending dependencies~(\S\ref{sec:illustration:cross-ecosystem}).
Packages which vendor packages from other ecosystems are a prime example, like the \texttt{polars} OCaml package with a number of Rust packages.
Another is packages which have some ad-hoc mechanism for invoking system packages managers to provide dependencies, such as opam's external dependencies mechanism.
We can mechanically translate external dependencies to cross-ecosystem dependencies for various operating systems.

We have written a custom resolver to operate on this hypergraph using a SAT encoding.
Despite the large number of packages, the performance of the resolution is a function of the size of the dependency cone and has remained tractable thus far.
Figure \ref{fig:solve} diagrams a subgraph of the resolved graph resulting from the query set containing opam's \texttt{polars.1.0.0} library described in Figure \ref{fig:packulus-miraculous}.
The Debian dependencies come from a dependency on the system Rust compiler (Figure~\ref{fig:solve} (a)), and the Cargo packages come from the Rust component of this bindings library.
Note the cycle between \texttt{deb-libgcc1.1-8.3.0-6} and \texttt{deb-libc6.2.28-10+deb10u1} which is broken by Debian~\cite{debian} (Figure~\ref{fig:solve} (b)), and the cycle between \texttt{opam-ocaml.4.14.3} and \texttt{opam-ocaml-variants.4.14.3+trunk} which is broken by a post variable (Figure~\ref{fig:solve} (c))~\cite{opam}.
Also note the use of optional dependencies which have been resolved to a concrete set of features encoded as dependencies of \texttt{opam-ocaml-variants.4.14.3+trunk}.
In the current state these Rust dependencies are vendored into the projects source repository, and Debian dependencies are ad-hoc and unversioned.

As every ecosystem has a different translation to the resolution hypergraph, so they all have a different mechanism of providing the resulting packages~(\S\ref{sec:pipeline:deployment}).
Some are easier to integrate than others.
We can perform a topological sort over the resolved graph and, for example, Debian's \texttt{dpkg} can be easily invoked to install a single package.
Alpine's APK can be used to install a package and resolve dependencies, but if we invoke it in topological order it will simply verify that our resolve has indeed provided all the required dependencies.
Even monolithic tools like Rust's Cargo can similarly be invoked to provide HyperRes resolved dependencies.
% Ryan: TODO how...?

\subsection{Ecosystem Translations}
\label{sec:ecosystem:translation}

So far, we discussed the $N+N$ problem for $N$ ecosystems: we write translations from ecosystem metadata to the resolution hypergraph ($N$) and interface with ecosystem tooling in order to deploy them ($N$).
This allows us to resolve dependencies across ecosystems and make package managers interoperable.
However, there is also an $N\times N$ problem of translating between ecosystem metadata bundles. This would allow the many millions of packages spread across ecosystems (Table~\ref{tbl:comparison}) to be directly incorporated into other ecosystems.
For example, Debian packages could be created directly from all the language-specific package managers that are binary compatible with the existing Debian packages, and without requiring the labour-intensive process of creating a Debian package for each language-specific package.

This translation process is easier for some package managers than others, and made significantly easier by the HyperRes formalism.
For example, translating an \text{opam} file to a \text{deb} archive is quite feasible due to the simple archive format of Debian source packages.
Others are harder, like the Nix package manager, but given an API into the Nix store the tools could create derivations \textit{translated} from another ecosystem.
Such metadata translations would allow the resolved graphs to all be translated to one target ecosystem,  allowing for the deployment of packages using a single tool.

Package managers are often platform-specific, but this `protocol' for package management would allow for package management systems specific to platforms, using technologies like namespaces on Linux and Jails on FreeBSD, still to interoperate.
This enables a new generation of package managers using a common protocol that can be hyper-specialised to an operating system or environment.

% common protocol with hyper-specialised managers for different systems
% Interoperability.
% Protocol for expression dependencies that can be implemented many ways.

\section{Related Work}
\label{sec:related}

\paragraph{SAT for Package Resolving.}
The core idea of using a formalism to aid with package versioning dates back to Opium~\cite{tucker2007opium} in 2007, which adapted Debian's apt-get to use a SAT solver for dependency resolution.
Opium showed that such resolving could be done efficiently, and later work on CUDF~\cite{cudf} specified a file format that could be used by specialised package resolvers.
One such early resolver used answer-set programming~\cite{gebser_et_al:LIPIcs.ICLP.2011.1}, which also found use in high-performance computing software infrastructure for solving dependencies with support for a rich set of user-specified resolver preferences~\cite{10046107}.
However, these systems did not halt the proliferation of package managers in the past decade, and the problem of cross-ecosystem dependencies has only grown worse in the intervening years.

Abate et al.~noted in 2020 that while SAT-based dependency solving did find traction among some package manager implementations, the idea of using reusable components for version solving had not taken off~\cite{9054837}.
In recent years, some of the larger package ecosystems such as NPM are under severe scaling pressure, and it is becoming increasingly important to revisit the question of how to reuse knowledge across ecosystems.
For example, PacSolve's MaxSMT based resolver is a drop-in replacement to NPM that reduced vulnerabilities in resolved packages and picked fresher and fewer dependencies~\cite{10172612}.
Approximate solving has also been useful for package selection under bounded latency situations~\cite{10.1145/2568225.2568306}.
However, these systems are still tied to a single ecosystem, and HyperRes is our attempt to bridge this gap and provide a `babelfish' for package management that will allow for the reuse of knowledge across ecosystems.
% https://dl.acm.org/doi/10.1016/j.infsof.2012.09.002
% https://www.mancoosi.org/papers/cbse11.pdf
% https://core.ac.uk/download/pdf/47091395.pdf

\paragraph{Open-source Ecosystems.}
There is an enormous and increasing amount of open-source software being published, and the problem of managing dependencies is only going to get worse.
Software Heritage is building a universal archive of this source code~\cite{10.1145/3183558}, and HyperRes offers the chance to build a higher-level and cross-ecosystem semantic representation of these interdependencies.

Maintainers of some of the key distributions such as Debian (which form the basis for many other downstream products) are having to scale their automation in order to manage the ever-growing package set~\cite{Becker2022} and vulnerabilities~\cite{Lin2023}.
The HyperRes approach of bringing common semantic meaning across language ecosystems should help reuse tooling that is currently language-specific to support the distribution maintainers such as those in Debian or CentOS.
Being able to reason about large-scale dependencies is also important to alleviating the reproducibility crisis in scientific computing~\cite{10.1145/3186266}, and HyperRes could (with sufficient tooling) support the longer-term availability of published scientific results.

\paragraph{Build Systems.}
Package management also has a close tie-in to build systems, which are equally surprisingly poorly-specified~\cite{mokhov2018build}.
Build systems and package management systems are sometimes unified in newer toolchains (Table~\ref{tbl:comparison}), but doing so makes it difficult to compose projects that span ecosystems (e.g.\ consider the difficulty of depending on a Rust library from OCaml code with just two package managers involved).
Therefore, the HyperRes approach of specifying a formalism that can be implemented as a translation library across package ecosystems maintains the separation from the underlying build systems.
In the long term, creating a composable formal theory of build systems and package managers~\cite{Agnarsson1985} would let us combine cross-ecosystem codebases to conduct correctness testing across the vast amount of source code published~\cite{MACHO2024111916}.
It would also allow the use of fuzzing techniques on HyperRes to spot incorrect packaging specifications, which is a technique successfully applied to build systems~\cite{8812082}.

\paragraph{Software Supply Chain Security.}
The increasing complexity of cross-ecosystem dependency chains puts pressure on the security of software supply chains, both on the choice of dependencies selected for a software project~\cite{Alfadel2023,10.1145/3571848,10298471} but also in vulnerabilities in the many package managers available~\cite{10.1145/1455770.1455841,arxiv.2302.08959}.
Software Bill of Materials (SBOMs) manifests are being rapidly adopted~\cite{arxiv.2409.01214} to specify the full set of versioned dependencies that go into a given application.
However, the same problem of uneven quality across ecosystems applies here, with the JavaScript ecosystem being particularly unpredictable~\cite{10.1145/3605098.3635927}.
HyperRes could be used to generate SBOMs systematically, to reduce transitive trust dependencies across language and OS ecosystems~\cite{schwaighofer2024extending}, drop resource usage by debloating unnecessary packages~\cite{10.1145/3488932.3524054}, and reduce the latency of applying security updates in third party dependencies~\cite{10.1007/978-3-031-21388-5_7,9359290}.

Some popular languages, such as C++, have had surprisingly limited uptake of package management.
Survey results~\cite{10.1145/3167132.3167290} show that many of the features needed by users to adopt package management are reflected in the HyperRes formalism set, but users also require close integration with the system package manager --- another new feature supported by HyperRes.

\paragraph{Human-in-the-loop Automation.}
LLMs are increasingly used for code generation tasks, and package management metadata is well-placed for automation here, with some early work showing promising results~\cite{DiazPace2024}.
One future direction that HyperRes could go in is to support LLM-driven human-in-the-loop package management~\cite{guo2024packageintel} via its concise and information-dense hypergraph representation.
Similar work on automating Dockerfile repair~\cite{9402069} has shown promising results, and the same techniques could apply to simplifying the housekeeping of the many package manager metadata files in use today.
%We present a more limited cost function than opium

%Performance discussion

%https://inria.hal.science/hal-00697463v1/document
%https://arxiv.org/pdf/2011.07851
%https://manpages.debian.org/unstable/alien/alien.1p.en.html
%https://github.com/mildred/alien

% https://www.mancoosi.org/reports/tr3.pdf
% without all the complexities of CUDF
% simpler
% mechanisable

%Note that in fig \ref{fig:pipeline} we assume a \textit{functional} deployment model.

\section{Conclusion}
\label{sec:conclusion}

We have presented HyperRes, a formalism that finds commonality among the many diverse package managers in wide use for software development today.
We have surveyed a representative set of package managers in use today and used HyperRes to translate individual packaging systems into our formalism, and then taken advantage of modern constraint solvers to answer versioning queries that span operating system and programming language ecosystems.
And we have also shown a path to developing a `babelfish' for package management that can use HyperRes to support the bidirectional translation of packages between ecosystems.
This in turn will allow for the reuse of packaging knowledge across open-source ecosystems, and greatly simplify and secure the increasingly complex software supply chain in use today.

%Future work: efficient solvers
%actually build the thing
%hyper-specialised package managers

%proper performance work
%SBOMs.
%consequences - dev envs highly tailored multi-language
%opam generate nix drv file without breaking invariants
%iot images
%multiple graphs, common subsets, bind mounts
%This enables a new generation of package managers using a common protocol that can be hyper-specialised to an operating system or environment.

%The current implementation is a prototype and may not cover all features of every package manager listed.
%Performance optimisations are ongoing, and the solver may have limitations with extremely large dependency graphs.

\bibliographystyle{ACM-Reference-Format}
\bibliography{reference}


\begin{thebibliography}{59}

%%% the Web address.

\ifx \showCODEN    \undefined \def \showCODEN     #1{\unskip}     \fi
\ifx \showISBNx    \undefined \def \showISBNx     #1{\unskip}     \fi
\ifx \showISBNxiii \undefined \def \showISBNxiii  #1{\unskip}     \fi
\ifx \showISSN     \undefined \def \showISSN      #1{\unskip}     \fi
\ifx \showLCCN     \undefined \def \showLCCN      #1{\unskip}     \fi
\ifx \shownote     \undefined \def \shownote      #1{#1}          \fi
\ifx \showarticletitle \undefined \def \showarticletitle #1{#1}   \fi
\ifx \showURL      \undefined \def \showURL       {\relax}        \fi
% invisible to TeX
\providecommand\bibfield[2]{#2}
\providecommand\bibinfo[2]{#2}
\providecommand\natexlab[1]{#1}
\providecommand\showeprint[2][]{arXiv:#2}

\bibitem[sla(2012)]%
        {slackware}
 \bibinfo{year}{2012}\natexlab{}.
\newblock \bibinfo{title}{Slackware Package Management}.
\newblock
\urldef\tempurl%
\url{https://docs.slackware.com/slackware:package_management}
\showURL{%
\tempurl}


\bibitem[opa(2013)]%
        {opam}
 \bibinfo{year}{2013}\natexlab{}.
\newblock \bibinfo{title}{OCaml Package Manager, The opam manual}.
\newblock
\urldef\tempurl%
\url{https://opam.ocaml.org/doc/Manual.html}
\showURL{%
\tempurl}


\bibitem[lsb(2015)]%
        {lsb}
 \bibinfo{year}{2015}\natexlab{}.
\newblock \bibinfo{title}{Linux Standard Base specification}.
\newblock
\urldef\tempurl%
\url{https://refspecs.linuxfoundation.org/lsb.shtml}
\showURL{%
\tempurl}


\bibitem[car(2024a)]%
        {cargo}
 \bibinfo{year}{2024}\natexlab{a}.
\newblock \bibinfo{title}{The Cargo Book}.
\newblock
\urldef\tempurl%
\url{https://doc.rust-lang.org/cargo/}
\showURL{%
\tempurl}


\bibitem[car(2024b)]%
        {cargoResolver}
 \bibinfo{year}{2024}\natexlab{b}.
\newblock \bibinfo{title}{Cargo Resolver}.
\newblock
\urldef\tempurl%
\url{https://github.com/rust-lang/cargo/blob/15fbd2f607d4defc87053b8b76bf5038f2483cf4/src/cargo/core/resolver/mod.rs}
\showURL{%
\tempurl}


\bibitem[deb(2024a)]%
        {debianVersions}
 \bibinfo{year}{2024}\natexlab{a}.
\newblock \bibinfo{title}{Debian Policy Manual: Chapter 5 - Control files and
  their fields}.
\newblock
\urldef\tempurl%
\url{https://www.debian.org/doc/debian-policy/ch-controlfields.html}
\showURL{%
\tempurl}


\bibitem[deb(2024b)]%
        {debian}
 \bibinfo{year}{2024}\natexlab{b}.
\newblock \bibinfo{title}{Debian Policy Manual: Chapter 7 - Declaring
  relationships between packages}.
\newblock
\urldef\tempurl%
\url{https://www.debian.org/doc/debian-policy/ch-relationships.html}
\showURL{%
\tempurl}


\bibitem[goM(2024)]%
        {goModules}
 \bibinfo{year}{2024}\natexlab{}.
\newblock \bibinfo{title}{Go Modules Reference}.
\newblock
\urldef\tempurl%
\url{https://go.dev/ref/mod}
\showURL{%
\tempurl}


\bibitem[has(2024)]%
        {haskell}
 \bibinfo{year}{2024}\natexlab{}.
\newblock \bibinfo{title}{The Haskell Cabal}.
\newblock
\urldef\tempurl%
\url{https://www.haskell.org/cabal/}
\showURL{%
\tempurl}


\bibitem[npm(2024)]%
        {npmScripts}
 \bibinfo{year}{2024}\natexlab{}.
\newblock \bibinfo{title}{npm Docs Scripts}.
\newblock
\urldef\tempurl%
\url{https://docs.npmjs.com/cli/v10/using-npm/scripts}
\showURL{%
\tempurl}


\bibitem[Abate et~al\mbox{.}(2020)]%
        {9054837}
\bibfield{author}{\bibinfo{person}{Pietro Abate}, \bibinfo{person}{Roberto
  Di~Cosmo}, \bibinfo{person}{Georgios Gousios}, {and} \bibinfo{person}{Stefano
  Zacchiroli}.} \bibinfo{year}{2020}\natexlab{}.
\newblock \showarticletitle{Dependency Solving Is Still Hard, but We Are
  Getting Better at It}. In \bibinfo{booktitle}{\emph{2020 IEEE 27th
  International Conference on Software Analysis, Evolution and Reengineering
  (SANER)}}. \bibinfo{pages}{547--551}.
\newblock
\href{https://doi.org/10.1109/SANER48275.2020.9054837}{doi:\nolinkurl{10.1109/SANER48275.2020.9054837}}


\bibitem[Abate et~al\mbox{.}(2011)]%
        {10.1145/2000229.2000255}
\bibfield{author}{\bibinfo{person}{Pietro Abate}, \bibinfo{person}{Roberto
  DiCosmo}, \bibinfo{person}{Ralf Treinen}, {and} \bibinfo{person}{Stefano
  Zacchiroli}.} \bibinfo{year}{2011}\natexlab{}.
\newblock \showarticletitle{MPM: a modular package manager}. In
  \bibinfo{booktitle}{\emph{Proceedings of the 14th International ACM Sigsoft
  Symposium on Component Based Software Engineering}} (Boulder, Colorado, USA)
  \emph{(\bibinfo{series}{CBSE '11})}. \bibinfo{publisher}{Association for
  Computing Machinery}, \bibinfo{address}{New York, NY, USA},
  \bibinfo{pages}{179–188}.
\newblock
\showISBNx{9781450307239}
\href{https://doi.org/10.1145/2000229.2000255}{doi:\nolinkurl{10.1145/2000229.2000255}}


\bibitem[Abramatic et~al\mbox{.}(2018)]%
        {10.1145/3183558}
\bibfield{author}{\bibinfo{person}{Jean-Fran\c{c}ois Abramatic},
  \bibinfo{person}{Roberto Di~Cosmo}, {and} \bibinfo{person}{Stefano
  Zacchiroli}.} \bibinfo{year}{2018}\natexlab{}.
\newblock \showarticletitle{Building the universal archive of source code}.
\newblock \bibinfo{journal}{\emph{Commun. ACM}} \bibinfo{volume}{61},
  \bibinfo{number}{10} (\bibinfo{date}{Sept.} \bibinfo{year}{2018}),
  \bibinfo{pages}{29–31}.
\newblock
\showISSN{0001-0782}
\href{https://doi.org/10.1145/3183558}{doi:\nolinkurl{10.1145/3183558}}


\bibitem[Agnarsson and Krishnamoorthy(1985)]%
        {Agnarsson1985}
\bibfield{author}{\bibinfo{person}{Snorri Agnarsson} {and}
  \bibinfo{person}{M.~S. Krishnamoorthy}.} \bibinfo{year}{1985}\natexlab{}.
\newblock \showarticletitle{Towards a theory of packages}.
\newblock \bibinfo{journal}{\emph{ACM SIGPLAN Notices}} \bibinfo{volume}{20},
  \bibinfo{number}{7} (\bibinfo{date}{June} \bibinfo{year}{1985}),
  \bibinfo{pages}{117–130}.
\newblock
\showISSN{1558-1160}
\href{https://doi.org/10.1145/17919.806833}{doi:\nolinkurl{10.1145/17919.806833}}


\bibitem[Alfadel et~al\mbox{.}(2023a)]%
        {Alfadel2023}
\bibfield{author}{\bibinfo{person}{Mahmoud Alfadel},
  \bibinfo{person}{Diego~Elias Costa}, {and} \bibinfo{person}{Emad Shihab}.}
  \bibinfo{year}{2023}\natexlab{a}.
\newblock \showarticletitle{Empirical analysis of security vulnerabilities in
  Python packages}.
\newblock \bibinfo{journal}{\emph{Empirical Software Engineering}}
  \bibinfo{volume}{28}, \bibinfo{number}{3} (\bibinfo{date}{March}
  \bibinfo{year}{2023}).
\newblock
\showISSN{1573-7616}
\href{https://doi.org/10.1007/s10664-022-10278-4}{doi:\nolinkurl{10.1007/s10664-022-10278-4}}


\bibitem[Alfadel et~al\mbox{.}(2023b)]%
        {10.1145/3571848}
\bibfield{author}{\bibinfo{person}{Mahmoud Alfadel},
  \bibinfo{person}{Diego~Elias Costa}, \bibinfo{person}{Emad Shihab}, {and}
  \bibinfo{person}{Bram Adams}.} \bibinfo{year}{2023}\natexlab{b}.
\newblock \showarticletitle{On the Discoverability of npm Vulnerabilities in
  Node.js Projects}.
\newblock \bibinfo{journal}{\emph{ACM Trans. Softw. Eng. Methodol.}}
  \bibinfo{volume}{32}, \bibinfo{number}{4}, Article \bibinfo{articleno}{91}
  (\bibinfo{date}{May} \bibinfo{year}{2023}), \bibinfo{numpages}{27}~pages.
\newblock
\showISSN{1049-331X}
\href{https://doi.org/10.1145/3571848}{doi:\nolinkurl{10.1145/3571848}}


\bibitem[Artho et~al\mbox{.}(2012)]%
        {6224274}
\bibfield{author}{\bibinfo{person}{Cyrille Artho}, \bibinfo{person}{Kuniyasu
  Suzaki}, \bibinfo{person}{Roberto Di~Cosmo}, \bibinfo{person}{Ralf Treinen},
  {and} \bibinfo{person}{Stefano Zacchiroli}.} \bibinfo{year}{2012}\natexlab{}.
\newblock \showarticletitle{Why do software packages conflict?}. In
  \bibinfo{booktitle}{\emph{2012 9th IEEE Working Conference on Mining Software
  Repositories (MSR)}}. \bibinfo{pages}{141--150}.
\newblock
\href{https://doi.org/10.1109/MSR.2012.6224274}{doi:\nolinkurl{10.1109/MSR.2012.6224274}}


\bibitem[Becker et~al\mbox{.}(2022)]%
        {Becker2022}
\bibfield{author}{\bibinfo{person}{Benedikt Becker}, \bibinfo{person}{Nicolas
  Jeannerod}, \bibinfo{person}{Claude Marché}, \bibinfo{person}{Yann
  Régis-Gianas}, \bibinfo{person}{Mihaela Sighireanu}, {and}
  \bibinfo{person}{Ralf Treinen}.} \bibinfo{year}{2022}\natexlab{}.
\newblock \showarticletitle{The CoLiS platform for the analysis of maintainer
  scripts in Debian software packages}.
\newblock \bibinfo{journal}{\emph{International Journal on Software Tools for
  Technology Transfer}} \bibinfo{volume}{24}, \bibinfo{number}{5}
  (\bibinfo{date}{Sept.} \bibinfo{year}{2022}), \bibinfo{pages}{717–733}.
\newblock
\showISSN{1433-2787}
\href{https://doi.org/10.1007/s10009-022-00671-1}{doi:\nolinkurl{10.1007/s10009-022-00671-1}}


\bibitem[Bentley(1986)]%
        {10.1145/6424.315691}
\bibfield{author}{\bibinfo{person}{Jon Bentley}.}
  \bibinfo{year}{1986}\natexlab{}.
\newblock \showarticletitle{Programming pearls: little languages}.
\newblock \bibinfo{journal}{\emph{Commun. ACM}} \bibinfo{volume}{29},
  \bibinfo{number}{8} (\bibinfo{date}{Aug.} \bibinfo{year}{1986}),
  \bibinfo{pages}{711–721}.
\newblock
\showISSN{0001-0782}
\href{https://doi.org/10.1145/6424.315691}{doi:\nolinkurl{10.1145/6424.315691}}


\bibitem[Berge(1970)]%
        {bergeHypergraphs}
\bibfield{author}{\bibinfo{person}{Claude Berge}.}
  \bibinfo{year}{1970}\natexlab{}.
\newblock \bibinfo{booktitle}{\emph{Graphes et hypergraphes}}.
\newblock \bibinfo{publisher}{Dunod}, \bibinfo{address}{Paris, France}.
\newblock
\showLCCN{75564040}


\bibitem[Bos(2023)]%
        {arxiv.2302.08959}
\bibfield{author}{\bibinfo{person}{Aarnav~M. Bos}.}
  \bibinfo{year}{2023}\natexlab{}.
\newblock \bibinfo{title}{A Review of Attacks Against Language-Based Package
  Managers}.
\newblock
\href{https://doi.org/10.48550/ARXIV.2302.08959}{doi:\nolinkurl{10.48550/ARXIV.2302.08959}}


\bibitem[Cappos et~al\mbox{.}(2008)]%
        {10.1145/1455770.1455841}
\bibfield{author}{\bibinfo{person}{Justin Cappos}, \bibinfo{person}{Justin
  Samuel}, \bibinfo{person}{Scott Baker}, {and} \bibinfo{person}{John~H.
  Hartman}.} \bibinfo{year}{2008}\natexlab{}.
\newblock \showarticletitle{A look in the mirror: attacks on package managers}.
  In \bibinfo{booktitle}{\emph{Proceedings of the 15th ACM Conference on
  Computer and Communications Security}} (Alexandria, Virginia, USA)
  \emph{(\bibinfo{series}{CCS '08})}. \bibinfo{publisher}{Association for
  Computing Machinery}, \bibinfo{address}{New York, NY, USA},
  \bibinfo{pages}{565–574}.
\newblock
\showISBNx{9781595938107}
\href{https://doi.org/10.1145/1455770.1455841}{doi:\nolinkurl{10.1145/1455770.1455841}}


\bibitem[Cofano et~al\mbox{.}(2024)]%
        {arxiv.2409.01214}
\bibfield{author}{\bibinfo{person}{Serena Cofano}, \bibinfo{person}{Giacomo
  Benedetti}, {and} \bibinfo{person}{Matteo Dell'Amico}.}
  \bibinfo{year}{2024}\natexlab{}.
\newblock \bibinfo{title}{SBOM Generation Tools in the Python Ecosystem: an
  In-Detail Analysis}.
\newblock
\href{https://doi.org/10.48550/ARXIV.2409.01214}{doi:\nolinkurl{10.48550/ARXIV.2409.01214}}


\bibitem[Cook(1971)]%
        {cookSAT}
\bibfield{author}{\bibinfo{person}{Stephen~A. Cook}.}
  \bibinfo{year}{1971}\natexlab{}.
\newblock \showarticletitle{The complexity of theorem-proving procedures}. In
  \bibinfo{booktitle}{\emph{Proceedings of the Third Annual ACM Symposium on
  Theory of Computing}} (Shaker Heights, Ohio, USA)
  \emph{(\bibinfo{series}{STOC '71})}. \bibinfo{publisher}{Association for
  Computing Machinery}, \bibinfo{address}{New York, NY, USA},
  \bibinfo{pages}{151–158}.
\newblock
\showISBNx{9781450374644}
\href{https://doi.org/10.1145/800157.805047}{doi:\nolinkurl{10.1145/800157.805047}}


\bibitem[Court{\`e}s(2013)]%
        {courtes2013functional}
\bibfield{author}{\bibinfo{person}{Ludovic Court{\`e}s}.}
  \bibinfo{year}{2013}\natexlab{}.
\newblock \showarticletitle{Functional package management with guix}.
\newblock \bibinfo{journal}{\emph{arXiv preprint arXiv:1305.4584}}
  (\bibinfo{year}{2013}).
\newblock


\bibitem[Cox(2018)]%
        {coxGoMVS2018}
\bibfield{author}{\bibinfo{person}{Russ Cox}.} \bibinfo{year}{2018}\natexlab{}.
\newblock \bibinfo{title}{Go \& Versioning: Minimal Version Selection}.
\newblock
\urldef\tempurl%
\url{https://research.swtch.com/vgo-mvs}
\showURL{%
\tempurl}


\bibitem[Diaz~Pace et~al\mbox{.}(2024)]%
        {DiazPace2024}
\bibfield{author}{\bibinfo{person}{Andres Diaz~Pace}, \bibinfo{person}{Antonela
  Tommasel}, {and} \bibinfo{person}{Hernan~Ceferino Vazquez}.}
  \bibinfo{year}{2024}\natexlab{}.
\newblock \showarticletitle{The JavaScript Package Selection Task: A
  Comparative Experiment Using an LLM-based Approach}.
\newblock \bibinfo{journal}{\emph{CLEI Electronic Journal}}
  \bibinfo{volume}{27}, \bibinfo{number}{2} (\bibinfo{date}{July}
  \bibinfo{year}{2024}).
\newblock
\showISSN{0717-5000}
\href{https://doi.org/10.19153/cleiej.27.2.4}{doi:\nolinkurl{10.19153/cleiej.27.2.4}}


\bibitem[Dolstra et~al\mbox{.}(2004)]%
        {dolstraNixSafePolicyFree2004}
\bibfield{author}{\bibinfo{person}{Eelco Dolstra}, \bibinfo{person}{Merijn de
  Jonge}, {and} \bibinfo{person}{Eelco Visser}.}
  \bibinfo{year}{2004}\natexlab{}.
\newblock \showarticletitle{Nix: A Safe and Policy-Free System for Software
  Deployment}. In \bibinfo{booktitle}{\emph{Proceedings of the 18th USENIX
  Conference on System Administration}} (Atlanta, GA)
  \emph{(\bibinfo{series}{LISA '04})}. \bibinfo{publisher}{USENIX Association},
  \bibinfo{address}{USA}, \bibinfo{pages}{79–92}.
\newblock


\bibitem[Dolstra and L\"{o}h(2008)]%
        {10.1145/1411204.1411255}
\bibfield{author}{\bibinfo{person}{Eelco Dolstra} {and} \bibinfo{person}{Andres
  L\"{o}h}.} \bibinfo{year}{2008}\natexlab{}.
\newblock \showarticletitle{NixOS: a purely functional Linux distribution}. In
  \bibinfo{booktitle}{\emph{Proceedings of the 13th ACM SIGPLAN International
  Conference on Functional Programming}} (Victoria, BC, Canada)
  \emph{(\bibinfo{series}{ICFP '08})}. \bibinfo{publisher}{Association for
  Computing Machinery}, \bibinfo{address}{New York, NY, USA},
  \bibinfo{pages}{367–378}.
\newblock
\showISBNx{9781595939197}
\href{https://doi.org/10.1145/1411204.1411255}{doi:\nolinkurl{10.1145/1411204.1411255}}


\bibitem[Gamblin et~al\mbox{.}(2022)]%
        {10046107}
\bibfield{author}{\bibinfo{person}{Todd Gamblin}, \bibinfo{person}{Massimiliano
  Culpo}, \bibinfo{person}{Gregory Becker}, {and} \bibinfo{person}{Sergei
  Shudler}.} \bibinfo{year}{2022}\natexlab{}.
\newblock \showarticletitle{Using Answer Set Programming for HPC Dependency
  Solving}. In \bibinfo{booktitle}{\emph{SC22: International Conference for
  High Performance Computing, Networking, Storage and Analysis}}.
  \bibinfo{publisher}{IEEE}, \bibinfo{pages}{1--15}.
\newblock
\href{https://doi.org/10.1109/SC41404.2022.00040}{doi:\nolinkurl{10.1109/SC41404.2022.00040}}


\bibitem[Gamblin et~al\mbox{.}(2015)]%
        {spack}
\bibfield{author}{\bibinfo{person}{Todd Gamblin}, \bibinfo{person}{Matthew
  LeGendre}, \bibinfo{person}{Michael~R. Collette}, \bibinfo{person}{Gregory~L.
  Lee}, \bibinfo{person}{Adam Moody}, \bibinfo{person}{Bronis~R. de Supinski},
  {and} \bibinfo{person}{Scott Futral}.} \bibinfo{year}{2015}\natexlab{}.
\newblock \showarticletitle{The Spack package manager: bringing order to HPC
  software chaos}. In \bibinfo{booktitle}{\emph{Proceedings of the
  International Conference for High Performance Computing, Networking, Storage
  and Analysis}} (Austin, Texas) \emph{(\bibinfo{series}{SC '15})}.
  \bibinfo{publisher}{Association for Computing Machinery},
  \bibinfo{address}{New York, NY, USA}, Article \bibinfo{articleno}{40},
  \bibinfo{numpages}{12}~pages.
\newblock
\showISBNx{9781450337236}
\href{https://doi.org/10.1145/2807591.2807623}{doi:\nolinkurl{10.1145/2807591.2807623}}


\bibitem[Gebser et~al\mbox{.}(2011)]%
        {gebser_et_al:LIPIcs.ICLP.2011.1}
\bibfield{author}{\bibinfo{person}{Martin Gebser}, \bibinfo{person}{Roland
  Kaminski}, \bibinfo{person}{Benjamin Kaufmann}, {and}
  \bibinfo{person}{Torsten Schaub}.} \bibinfo{year}{2011}\natexlab{}.
\newblock \showarticletitle{{Multi-Criteria Optimization in Answer Set
  Programming}}. In \bibinfo{booktitle}{\emph{Technical Communications of the
  27th International Conference on Logic Programming (ICLP'11)}}
  \emph{(\bibinfo{series}{Leibniz International Proceedings in Informatics
  (LIPIcs)}, Vol.~\bibinfo{volume}{11})},
  \bibfield{editor}{\bibinfo{person}{John~P. Gallagher} {and}
  \bibinfo{person}{Michael Gelfond}} (Eds.). \bibinfo{publisher}{Schloss
  Dagstuhl -- Leibniz-Zentrum f{\"u}r Informatik}, \bibinfo{address}{Dagstuhl,
  Germany}, \bibinfo{pages}{1--10}.
\newblock
\showISBNx{978-3-939897-31-6}
\showISSN{1868-8969}
\href{https://doi.org/10.4230/LIPIcs.ICLP.2011.1}{doi:\nolinkurl{10.4230/LIPIcs.ICLP.2011.1}}


\bibitem[Guo et~al\mbox{.}(2024)]%
        {guo2024packageintel}
\bibfield{author}{\bibinfo{person}{Wenbo Guo}, \bibinfo{person}{Chengwei Liu},
  \bibinfo{person}{Limin Wang}, \bibinfo{person}{Jiahui Wu},
  \bibinfo{person}{Zhengzi Xu}, \bibinfo{person}{Cheng Huang},
  \bibinfo{person}{Yong Fang}, {and} \bibinfo{person}{Yang Liu}.}
  \bibinfo{year}{2024}\natexlab{}.
\newblock \showarticletitle{PackageIntel: Leveraging Large Language Models for
  Automated Intelligence Extraction in Package Ecosystems}.
\newblock \bibinfo{journal}{\emph{arXiv preprint arXiv:2409.15049}}
  (\bibinfo{year}{2024}).
\newblock


\bibitem[Henkel et~al\mbox{.}(2021)]%
        {9402069}
\bibfield{author}{\bibinfo{person}{Jordan Henkel}, \bibinfo{person}{Denini
  Silva}, \bibinfo{person}{Leopoldo Teixeira}, \bibinfo{person}{Marcelo
  d’Amorim}, {and} \bibinfo{person}{Thomas Reps}.}
  \bibinfo{year}{2021}\natexlab{}.
\newblock \showarticletitle{Shipwright: A Human-in-the-Loop System for
  Dockerfile Repair}. In \bibinfo{booktitle}{\emph{2021 IEEE/ACM 43rd
  International Conference on Software Engineering (ICSE)}}.
  \bibinfo{pages}{1148--1160}.
\newblock
\href{https://doi.org/10.1109/ICSE43902.2021.00106}{doi:\nolinkurl{10.1109/ICSE43902.2021.00106}}


\bibitem[Hudak(1996)]%
        {10.1145/242224.242477}
\bibfield{author}{\bibinfo{person}{Paul Hudak}.}
  \bibinfo{year}{1996}\natexlab{}.
\newblock \showarticletitle{Building domain-specific embedded languages}.
\newblock \bibinfo{journal}{\emph{ACM Comput. Surv.}} \bibinfo{volume}{28},
  \bibinfo{number}{4es} (\bibinfo{date}{Dec.} \bibinfo{year}{1996}),
  \bibinfo{pages}{196–es}.
\newblock
\showISSN{0360-0300}
\href{https://doi.org/10.1145/242224.242477}{doi:\nolinkurl{10.1145/242224.242477}}


\bibitem[Ignatiev et~al\mbox{.}(2014)]%
        {10.1145/2568225.2568306}
\bibfield{author}{\bibinfo{person}{Alexey Ignatiev},
  \bibinfo{person}{Mikol\'{a}\v{s} Janota}, {and} \bibinfo{person}{Joao
  Marques-Silva}.} \bibinfo{year}{2014}\natexlab{}.
\newblock \showarticletitle{Towards efficient optimization in package
  management systems}. In \bibinfo{booktitle}{\emph{Proceedings of the 36th
  International Conference on Software Engineering}} (Hyderabad, India)
  \emph{(\bibinfo{series}{ICSE 2014})}. \bibinfo{publisher}{Association for
  Computing Machinery}, \bibinfo{address}{New York, NY, USA},
  \bibinfo{pages}{745–755}.
\newblock
\showISBNx{9781450327565}
\href{https://doi.org/10.1145/2568225.2568306}{doi:\nolinkurl{10.1145/2568225.2568306}}


\bibitem[Ivie and Thain(2018)]%
        {10.1145/3186266}
\bibfield{author}{\bibinfo{person}{Peter Ivie} {and} \bibinfo{person}{Douglas
  Thain}.} \bibinfo{year}{2018}\natexlab{}.
\newblock \showarticletitle{Reproducibility in Scientific Computing}.
\newblock \bibinfo{journal}{\emph{ACM Comput. Surv.}} \bibinfo{volume}{51},
  \bibinfo{number}{3}, Article \bibinfo{articleno}{63} (\bibinfo{date}{July}
  \bibinfo{year}{2018}), \bibinfo{numpages}{36}~pages.
\newblock
\showISSN{0360-0300}
\href{https://doi.org/10.1145/3186266}{doi:\nolinkurl{10.1145/3186266}}


\bibitem[Lamb and Zacchiroli(2021)]%
        {reproducibleBuilds}
\bibfield{author}{\bibinfo{person}{Chris Lamb} {and} \bibinfo{person}{Stefano
  Zacchiroli}.} \bibinfo{year}{2021}\natexlab{}.
\newblock \showarticletitle{Reproducible Builds: Increasing the Integrity of
  Software Supply Chains}.
\newblock \bibinfo{journal}{\emph{CoRR}}  \bibinfo{volume}{abs/2104.06020}
  (\bibinfo{year}{2021}).
\newblock
\showeprint[arXiv]{2104.06020}
\urldef\tempurl%
\url{https://arxiv.org/abs/2104.06020}
\showURL{%
\tempurl}


\bibitem[Licker and Rice(2019)]%
        {8812082}
\bibfield{author}{\bibinfo{person}{Nandor Licker} {and} \bibinfo{person}{Andrew
  Rice}.} \bibinfo{year}{2019}\natexlab{}.
\newblock \showarticletitle{Detecting Incorrect Build Rules}. In
  \bibinfo{booktitle}{\emph{2019 IEEE/ACM 41st International Conference on
  Software Engineering (ICSE)}}. \bibinfo{pages}{1234--1244}.
\newblock
\href{https://doi.org/10.1109/ICSE.2019.00125}{doi:\nolinkurl{10.1109/ICSE.2019.00125}}


\bibitem[Lin et~al\mbox{.}(2023)]%
        {Lin2023}
\bibfield{author}{\bibinfo{person}{Jiahuei Lin}, \bibinfo{person}{Haoxiang
  Zhang}, \bibinfo{person}{Bram Adams}, {and} \bibinfo{person}{Ahmed~E.
  Hassan}.} \bibinfo{year}{2023}\natexlab{}.
\newblock \showarticletitle{Vulnerability management in Linux distributions: An
  empirical study on Debian and Fedora}.
\newblock \bibinfo{journal}{\emph{Empirical Software Engineering}}
  \bibinfo{volume}{28}, \bibinfo{number}{2} (\bibinfo{date}{Feb.}
  \bibinfo{year}{2023}).
\newblock
\showISSN{1573-7616}
\href{https://doi.org/10.1007/s10664-022-10267-7}{doi:\nolinkurl{10.1007/s10664-022-10267-7}}


\bibitem[Macho et~al\mbox{.}(2024)]%
        {MACHO2024111916}
\bibfield{author}{\bibinfo{person}{Christian Macho}, \bibinfo{person}{Fabian
  Oraze}, {and} \bibinfo{person}{Martin Pinzger}.}
  \bibinfo{year}{2024}\natexlab{}.
\newblock \showarticletitle{DValidator: An approach for validating dependencies
  in build configurations}.
\newblock \bibinfo{journal}{\emph{Journal of Systems and Software}}
  \bibinfo{volume}{209} (\bibinfo{year}{2024}), \bibinfo{pages}{111916}.
\newblock
\showISSN{0164-1212}
\href{https://doi.org/10.1016/j.jss.2023.111916}{doi:\nolinkurl{10.1016/j.jss.2023.111916}}


\bibitem[Mancinelli et~al\mbox{.}(2006)]%
        {4019575}
\bibfield{author}{\bibinfo{person}{Fabio Mancinelli}, \bibinfo{person}{Jaap
  Boender}, \bibinfo{person}{Roberto di Cosmo}, \bibinfo{person}{Jerome
  Vouillon}, \bibinfo{person}{Berke Durak}, \bibinfo{person}{Xavier Leroy},
  {and} \bibinfo{person}{Ralf Treinen}.} \bibinfo{year}{2006}\natexlab{}.
\newblock \showarticletitle{Managing the Complexity of Large Free and Open
  Source Package-Based Software Distributions}. In
  \bibinfo{booktitle}{\emph{21st IEEE/ACM International Conference on Automated
  Software Engineering (ASE'06)}}. \bibinfo{pages}{199--208}.
\newblock
\href{https://doi.org/10.1109/ASE.2006.49}{doi:\nolinkurl{10.1109/ASE.2006.49}}


\bibitem[Miranda and Pimentel(2018)]%
        {10.1145/3167132.3167290}
\bibfield{author}{\bibinfo{person}{Andr\'{e} Miranda} {and}
  \bibinfo{person}{Jo\~{a}o Pimentel}.} \bibinfo{year}{2018}\natexlab{}.
\newblock \showarticletitle{On the use of package managers by the C++
  open-source community}. In \bibinfo{booktitle}{\emph{Proceedings of the 33rd
  Annual ACM Symposium on Applied Computing}} (Pau, France)
  \emph{(\bibinfo{series}{SAC '18})}. \bibinfo{publisher}{Association for
  Computing Machinery}, \bibinfo{address}{New York, NY, USA},
  \bibinfo{pages}{1483–1491}.
\newblock
\showISBNx{9781450351911}
\href{https://doi.org/10.1145/3167132.3167290}{doi:\nolinkurl{10.1145/3167132.3167290}}


\bibitem[Mokhov et~al\mbox{.}(2018)]%
        {mokhov2018build}
\bibfield{author}{\bibinfo{person}{Andrey Mokhov}, \bibinfo{person}{Neil
  Mitchell}, {and} \bibinfo{person}{Simon Peyton~Jones}.}
  \bibinfo{year}{2018}\natexlab{}.
\newblock \showarticletitle{Build systems a la carte}. In
  \bibinfo{booktitle}{\emph{International Conference on Functional Programming
  (ICFP'18)}}. \bibinfo{publisher}{ACM}.
\newblock
\urldef\tempurl%
\url{https://www.microsoft.com/en-us/research/publication/build-systems-la-carte/}
\showURL{%
\tempurl}


\bibitem[Muhammad et~al\mbox{.}(2019)]%
        {10.1145/3365137.3365402}
\bibfield{author}{\bibinfo{person}{Hisham Muhammad}, \bibinfo{person}{Lucas
  C.~Villa Real}, {and} \bibinfo{person}{Michael Homer}.}
  \bibinfo{year}{2019}\natexlab{}.
\newblock \showarticletitle{Taxonomy of Package Management in Programming
  Languages and Operating Systems}. In \bibinfo{booktitle}{\emph{Proceedings of
  the 10th Workshop on Programming Languages and Operating Systems}}
  (Huntsville, ON, Canada) \emph{(\bibinfo{series}{PLOS '19})}.
  \bibinfo{publisher}{Association for Computing Machinery},
  \bibinfo{address}{New York, NY, USA}, \bibinfo{pages}{60–66}.
\newblock
\showISBNx{9781450370172}
\href{https://doi.org/10.1145/3365137.3365402}{doi:\nolinkurl{10.1145/3365137.3365402}}


\bibitem[Openja et~al\mbox{.}(2022)]%
        {10.1145/3530019.3530039}
\bibfield{author}{\bibinfo{person}{Moses Openja}, \bibinfo{person}{Forough
  Majidi}, \bibinfo{person}{Foutse Khomh}, \bibinfo{person}{Bhagya
  Chembakottu}, {and} \bibinfo{person}{Heng Li}.}
  \bibinfo{year}{2022}\natexlab{}.
\newblock \showarticletitle{Studying the Practices of Deploying Machine
  Learning Projects on Docker}. In \bibinfo{booktitle}{\emph{Proceedings of the
  26th International Conference on Evaluation and Assessment in Software
  Engineering}} (Gothenburg, Sweden) \emph{(\bibinfo{series}{EASE '22})}.
  \bibinfo{publisher}{Association for Computing Machinery},
  \bibinfo{address}{New York, NY, USA}, \bibinfo{pages}{190–200}.
\newblock
\showISBNx{9781450396134}
\href{https://doi.org/10.1145/3530019.3530039}{doi:\nolinkurl{10.1145/3530019.3530039}}


\bibitem[Pashakhanloo et~al\mbox{.}(2022)]%
        {10.1145/3488932.3524054}
\bibfield{author}{\bibinfo{person}{Pardis Pashakhanloo},
  \bibinfo{person}{Aravind Machiry}, \bibinfo{person}{Hyonyoung Choi},
  \bibinfo{person}{Anthony Canino}, \bibinfo{person}{Kihong Heo},
  \bibinfo{person}{Insup Lee}, {and} \bibinfo{person}{Mayur Naik}.}
  \bibinfo{year}{2022}\natexlab{}.
\newblock \showarticletitle{PacJam: Securing Dependencies Continuously via
  Package-Oriented Debloating}. In \bibinfo{booktitle}{\emph{Proceedings of the
  2022 ACM on Asia Conference on Computer and Communications Security}}
  (Nagasaki, Japan) \emph{(\bibinfo{series}{ASIA CCS '22})}.
  \bibinfo{publisher}{Association for Computing Machinery},
  \bibinfo{address}{New York, NY, USA}, \bibinfo{pages}{903–916}.
\newblock
\showISBNx{9781450391405}
\href{https://doi.org/10.1145/3488932.3524054}{doi:\nolinkurl{10.1145/3488932.3524054}}


\bibitem[Pinckney et~al\mbox{.}(2023)]%
        {10172612}
\bibfield{author}{\bibinfo{person}{Donald Pinckney}, \bibinfo{person}{Federico
  Cassano}, \bibinfo{person}{Arjun Guha}, \bibinfo{person}{Jonathan Bell},
  \bibinfo{person}{Massimiliano Culpo}, {and} \bibinfo{person}{Todd Gamblin}.}
  \bibinfo{year}{2023}\natexlab{}.
\newblock \showarticletitle{Flexible and Optimal Dependency Management via
  Max-SMT}. In \bibinfo{booktitle}{\emph{2023 IEEE/ACM 45th International
  Conference on Software Engineering (ICSE)}}. \bibinfo{publisher}{IEEE},
  \bibinfo{pages}{1418--1429}.
\newblock
\href{https://doi.org/10.1109/ICSE48619.2023.00124}{doi:\nolinkurl{10.1109/ICSE48619.2023.00124}}


\bibitem[Preston-Werner(2023)]%
        {semver}
\bibfield{author}{\bibinfo{person}{Tom Preston-Werner}.}
  \bibinfo{year}{2023}\natexlab{}.
\newblock \bibinfo{title}{{Semantic Versioning 2.0.0}}.
\newblock
\urldef\tempurl%
\url{https://semver.org/}
\showURL{%
\tempurl}


\bibitem[Primiero and Boender(2018)]%
        {primiero2018negative}
\bibfield{author}{\bibinfo{person}{Giuseppe Primiero} {and}
  \bibinfo{person}{Jaap Boender}.} \bibinfo{year}{2018}\natexlab{}.
\newblock \showarticletitle{Negative trust for conflict resolution in software
  management}. In \bibinfo{booktitle}{\emph{Web Intelligence}},
  Vol.~\bibinfo{volume}{16}. IOS Press, \bibinfo{pages}{251--271}.
\newblock


\bibitem[Rabbi et~al\mbox{.}(2024)]%
        {10.1145/3605098.3635927}
\bibfield{author}{\bibinfo{person}{Md~Fazle Rabbi},
  \bibinfo{person}{Arifa~Islam Champa}, \bibinfo{person}{Costain Nachuma},
  {and} \bibinfo{person}{Minhaz~Fahim Zibran}.}
  \bibinfo{year}{2024}\natexlab{}.
\newblock \showarticletitle{SBOM Generation Tools Under Microscope: A Focus on
  The npm Ecosystem}. In \bibinfo{booktitle}{\emph{Proceedings of the 39th
  ACM/SIGAPP Symposium on Applied Computing}} (Avila, Spain)
  \emph{(\bibinfo{series}{SAC '24})}. \bibinfo{publisher}{Association for
  Computing Machinery}, \bibinfo{address}{New York, NY, USA},
  \bibinfo{pages}{1233–1241}.
\newblock
\showISBNx{9798400702433}
\href{https://doi.org/10.1145/3605098.3635927}{doi:\nolinkurl{10.1145/3605098.3635927}}


\bibitem[Rahkema and Pfahl(2022)]%
        {10.1007/978-3-031-21388-5_7}
\bibfield{author}{\bibinfo{person}{Kristiina Rahkema} {and}
  \bibinfo{person}{Dietmar Pfahl}.} \bibinfo{year}{2022}\natexlab{}.
\newblock \showarticletitle{Analysing the Relationship Between Dependency
  Definition and Updating Practice When Using Third-Party Libraries}. In
  \bibinfo{booktitle}{\emph{Product-Focused Software Process Improvement}},
  \bibfield{editor}{\bibinfo{person}{Davide Taibi}, \bibinfo{person}{Marco
  Kuhrmann}, \bibinfo{person}{Tommi Mikkonen}, \bibinfo{person}{Jil
  Kl{\"u}nder}, {and} \bibinfo{person}{Pekka Abrahamsson}} (Eds.).
  \bibinfo{publisher}{Springer International Publishing},
  \bibinfo{address}{Cham}, \bibinfo{pages}{90--107}.
\newblock
\showISBNx{978-3-031-21388-5}


\bibitem[Schwaighofer et~al\mbox{.}(2024)]%
        {schwaighofer2024extending}
\bibfield{author}{\bibinfo{person}{Martin Schwaighofer},
  \bibinfo{person}{Michael Roland}, {and} \bibinfo{person}{Ren{\'e}
  Mayrhofer}.} \bibinfo{year}{2024}\natexlab{}.
\newblock \showarticletitle{Extending Cloud Build Systems to Eliminate
  Transitive Trust}. In \bibinfo{booktitle}{\emph{ACM Workshop on Software
  Supply Chain Offensive Research and Ecosystem Defenses (SCORED24)}} (Salt
  Lake City, Utah, USA). \bibinfo{publisher}{ACM}.
\newblock


\bibitem[Serrano and Gallesio(2007)]%
        {10.1145/1297081.1297093}
\bibfield{author}{\bibinfo{person}{Manuel Serrano} {and} \bibinfo{person}{Erick
  Gallesio}.} \bibinfo{year}{2007}\natexlab{}.
\newblock \showarticletitle{An adaptive package management system for scheme}.
  In \bibinfo{booktitle}{\emph{Proceedings of the 2007 Symposium on Dynamic
  Languages}} (Montreal, Quebec, Canada) \emph{(\bibinfo{series}{DLS '07})}.
  \bibinfo{publisher}{Association for Computing Machinery},
  \bibinfo{address}{New York, NY, USA}, \bibinfo{pages}{65–76}.
\newblock
\showISBNx{9781595938688}
\href{https://doi.org/10.1145/1297081.1297093}{doi:\nolinkurl{10.1145/1297081.1297093}}


\bibitem[Stringer et~al\mbox{.}(2020)]%
        {9359290}
\bibfield{author}{\bibinfo{person}{Jacob Stringer}, \bibinfo{person}{Amjed
  Tahir}, \bibinfo{person}{Kelly Blincoe}, {and} \bibinfo{person}{Jens
  Dietrich}.} \bibinfo{year}{2020}\natexlab{}.
\newblock \showarticletitle{Technical Lag of Dependencies in Major Package
  Managers}. In \bibinfo{booktitle}{\emph{2020 27th Asia-Pacific Software
  Engineering Conference (APSEC)}}. \bibinfo{pages}{228--237}.
\newblock
\href{https://doi.org/10.1109/APSEC51365.2020.00031}{doi:\nolinkurl{10.1109/APSEC51365.2020.00031}}


\bibitem[Treinen and Zacchiroli(2009)]%
        {cudf}
\bibfield{author}{\bibinfo{person}{Ralf Treinen} {and} \bibinfo{person}{Stefano
  Zacchiroli}.} \bibinfo{year}{2009}\natexlab{}.
\newblock \showarticletitle{Common Upgradeability Description Format (CUDF)
  2.0}.
\newblock \bibinfo{journal}{\emph{The Mancoosi project (FP7)}}
  \bibinfo{volume}{3} (\bibinfo{year}{2009}).
\newblock
\urldef\tempurl%
\url{https://www.mancoosi.org/reports/tr3.pdf}
\showURL{%
\tempurl}


\bibitem[Trezentos et~al\mbox{.}(2010)]%
        {10.1145/1858996.1859087}
\bibfield{author}{\bibinfo{person}{Paulo Trezentos}, \bibinfo{person}{In\^{e}s
  Lynce}, {and} \bibinfo{person}{Arlindo~L. Oliveira}.}
  \bibinfo{year}{2010}\natexlab{}.
\newblock \showarticletitle{Apt-pbo: solving the software dependency problem
  using pseudo-boolean optimization}. In \bibinfo{booktitle}{\emph{Proceedings
  of the 25th IEEE/ACM International Conference on Automated Software
  Engineering}} (Antwerp, Belgium) \emph{(\bibinfo{series}{ASE '10})}.
  \bibinfo{publisher}{Association for Computing Machinery},
  \bibinfo{address}{New York, NY, USA}, \bibinfo{pages}{427–436}.
\newblock
\showISBNx{9781450301169}
\href{https://doi.org/10.1145/1858996.1859087}{doi:\nolinkurl{10.1145/1858996.1859087}}


\bibitem[Tucker et~al\mbox{.}(2007)]%
        {tucker2007opium}
\bibfield{author}{\bibinfo{person}{Chris Tucker}, \bibinfo{person}{David
  Shuffelton}, \bibinfo{person}{Ranjit Jhala}, {and} \bibinfo{person}{Sorin
  Lerner}.} \bibinfo{year}{2007}\natexlab{}.
\newblock \showarticletitle{Opium: Optimal package install/uninstall manager}.
  In \bibinfo{booktitle}{\emph{29th International Conference on Software
  Engineering (ICSE'07)}}. IEEE, \bibinfo{publisher}{ACM},
  \bibinfo{address}{Minneapolis, MN, USA}, \bibinfo{pages}{178--188}.
\newblock


\bibitem[Zhang et~al\mbox{.}(2023)]%
        {10298471}
\bibfield{author}{\bibinfo{person}{Lyuye Zhang}, \bibinfo{person}{Chengwei
  Liu}, \bibinfo{person}{Sen Chen}, \bibinfo{person}{Zhengzi Xu},
  \bibinfo{person}{Lingling Fan}, \bibinfo{person}{Lida Zhao},
  \bibinfo{person}{Yiran Zhang}, {and} \bibinfo{person}{Yang Liu}.}
  \bibinfo{year}{2023}\natexlab{}.
\newblock \showarticletitle{Mitigating Persistence of Open-Source
  Vulnerabilities in Maven Ecosystem}. In \bibinfo{booktitle}{\emph{2023 38th
  IEEE/ACM International Conference on Automated Software Engineering (ASE)}}.
  \bibinfo{publisher}{ACM/IEEE}, \bibinfo{pages}{191--203}.
\newblock
\href{https://doi.org/10.1109/ASE56229.2023.00058}{doi:\nolinkurl{10.1109/ASE56229.2023.00058}}


\end{thebibliography}

\end{document}